\documentclass[11pt,numbers=enddot,a4paper]{scrartcl}

\usepackage{amsthm,amssymb,amsmath}
\usepackage{wasysym,latexsym,bm}
\usepackage{graphicx,psfrag,color}
\usepackage[ansinew]{inputenc}
\usepackage[USenglish]{babel}

\usepackage{fourier} 
\usepackage[scaled=0.875]{helvet} 
\usepackage{microtype} 
\usepackage{lettrine} 
\usepackage{paralist} 
\usepackage{abstract} 
\usepackage{booktabs}
\usepackage[flushmargin,multiple]{footmisc}
\usepackage[comma]{natbib}
\usepackage[top=3.5cm,textheight=23.5cm,textwidth=15cm,footskip=1.5cm]{geometry} 
\usepackage{fancyhdr} 
\newfont{\euro}{eurorm scaled \magstephalf}  

\renewcommand{\d}{\textup{d}}

\renewcommand{\P}{\mathbb{P}}
\newcommand{\Q}{\mathbb{Q}}

\newcommand{\1}{\mathbb{1}}
\newcommand{\N}{\mathbb{N}}

\newcommand{\R}{\mathbb{R}}

\newcommand{\E}{\textup{E}}

\newcommand{\QED}{\hfill Q.E.D.\medskip}

\newtheorem{theo}{Theorem}
\newtheorem{prop}{Proposition}
\newtheorem{lem}{Lemma}

\newtheorem{defn}{Definition}
\theoremstyle{definition}

\setkomafont{title}{\rm}%
\setkomafont{section}{\Large\centering\rm}%
\setkomafont{subsection}{\large\centering\rm}%
\setkomafont{subsubsection}{\centering\rm}

\begin{document}
\title{\textbf{Pricing and Valuation under the \\
Real-World Measure}}
\author{Gabriel Frahm\thanks{\hspace{1ex}Phone: +49 40 6541-2791, e-mail: frahm@hsu-hh.de.\vspace{.5em}}\\
Helmut Schmidt University\\
Department of Mathematics/Statistics\\
Chair for Applied Stochastics and \\
Risk Management}


\maketitle
\thispagestyle{fancy}

\begin{abstract}
\noindent In general it is not clear which kind of information is supposed to be used for calculating the fair value of a contingent claim. Even if the information is specified, it is not guaranteed that the fair value is uniquely determined by the given information. A further problem is that asset prices are typically expressed in terms of a risk-neutral measure. This makes it difficult to transfer the fundamental results of financial mathematics to econometrics. I show that the aforementioned problems evaporate if the financial market is complete and sensitive. In this case, after an appropriate choice of the numéraire, the discounted price processes turn out to be uniformly integrable martingales under the real-world measure. This leads to a Law of One Price and a simple real-world valuation formula in a model-independent framework where the number of assets as well as the lifetime of the market can be finite or infinite.
\end{abstract}

\smallskip


\newpage

\tableofcontents

\section{Motivation}

\lettrine[lines=3,nindent=0pt]{T}{he} central motivation of this work is to clarify the economic conditions under which the discounted price processes in a financial market are martingales under the \emph{physical} measure $\P$ and not only under an equivalent martingale measure $\Q\neq\P$. This martingale property is strongly connected to Samuelson's Martingale Hypothesis, which is also formulated in terms of $\P$ instead of $\Q$ \citep{Samuelson1965}. A substantial difference between Samuelson's approach and the methodological framework chosen in this work is that the desired martingale property is derived without any requirement on the interest and risk attitude of the market participants. The underlying probabilistic assumptions are minimal. In this model-independent framework, I try to build a bridge between the fundamental results of financial mathematics in terms of ``$\Q$'' and the broad field of econometrics, which requires the ``$\P$.''

Let $\textbf{F}$ be any flow of information that encompasses the evolution $\textbf{E}$ of asset prices in a complete financial market. The main result of this work can be stated as follows:
\begin{quote}
\textsl{If the market is sensitive to $\textbf{F}$, there exists a normalized $\textbf{E}$-predictable trading strategy that can be chosen as a numéraire such that each discounted price process is a uniformly integrable $\P$-martingale with respect to $\textbf{F}$.}
\end{quote}
\begin{quote}
\textsl{Conversely, choose any normalized $\textbf{E}$-predictable trading strategy as a numéraire. If each discounted price process is a uniformly integrable $\P$-martingale with respect to $\textbf{F}$, the market is sensitive to $\textbf{F}$.}
\end{quote}
\begin{quote}
\textsl{In either case, the chosen numéraire is the unique growth-optimal portfolio with respect to $\textbf{F}$, and $\P$ is the unique equivalent measure under which the discounted price process is a uniformly integrable martingale with respect to $\textbf{F}$.}
\end{quote}

In the following, every financial market is said to be \textit{simple} if and only if it contains a finite number of assets. By contrast, it is said to be \textit{complex} if and only if the number of assets is infinite. The main result solves a fundamental problem which frequently occurs in the context of pricing and valuation both in simple and complex financial markets. This problem is threefold:\label{T.: Problems (verbal)}
\begin{enumerate}
\item[(i)] The set of equivalent martingale measures depends on the given information flow. Hence, there are many possibilities to represent the asset prices and to calculate the fair value of a contingent claim. This leads to the following question:
    \begin{quote}
    \textsl{Does it pay to strive for more information or is it better to renounce searching altogether and to use the information we already have?}
    \end{quote}
\item[(ii)] Even if we specify the flow of information, the set of equivalent martingale measures typically contains a multitude of elements. In this case, it is still not clear which one to choose and then the fair value of a contingent claim is not uniquely determined by the given information. Hence, we might ask:
    \begin{quote}
    \textsl{Which economic condition guarantees that the set of equivalent martingale measures is a singleton given the specified flow of information?}
    \end{quote}
\item[(iii)] Given a unique equivalent martingale measure for the specified flow of information, it is not always clear how to use this measure in empirical applications, especially if the market is complex. Therefore, the last question is:
    \begin{quote}
    \textsl{Under which circumstances is it possible to represent asset prices and calculate the fair value of any contingent claim in terms of\/ $\P$ instead of\/ $\Q$?}
    \end{quote}
\end{enumerate}

These issues are highly relevant both from a theoretical and a practical perspective. Albeit the given exposition is rigorous in a mathematical sense, most of the presented results have a clear economic content. In particular, the results developed in this work fit harmonically into different coexistent branches of financial mathematics and finance theory. I hope that their practical implications are substantial. The industry still keeps inventing complicated financial instruments, which is a permanent challenge for the quant. This work shall provide a universal approach for assessing the fair value of a contingent claim, which might be considered helpful for the practitioner.

The main result of this work requires a complete financial market. Unfortunately, the classic notion of market completeness has got a bad reputation. In simple financial markets, i.e., if the number of assets is finite, the assumption of market completeness is very restrictive. In the continuous-time framework, only a small number of models are known to be complete, e.g., Bachelier's Brownian-motion model, the Black-Scholes model, the compensated Poisson process, and Azéma martingales \citep{CR1976,HP1981,JP2008}. For this reason, many alternative approaches have been proposed during the last decades. In particular, the concept of market completeness has been adopted to complex financial markets, i.e., to markets with an infinite number of assets \citep[see, e.g.,][]{AH1995,BJ1999,Delbaen1992,JM1999,JJM1999}.\footnote{For a nice overview of those contributions see \citet{Biagini2010}.} On the one hand, this essentially relaxes the notion of market completeness, but on the other hand market complexity sets higher standards for the underlying economy. In view of the vast amount of financial instruments and the increasing globalization of financial markets, complexity can be regarded as an acceptable assumption, at least for every well-developed economy.

Similarly, one can find a plethora of definitions of market efficiency \citep[see, e.g.,][]{Fama1965,Fama1970,FFJR1969,Latham1986,Malkiel1992,Samuelson1965}. The classic approach to the Efficient-Market Hypothesis is based on the fair-game model \citep{Fama1970}. Unfortunately, this model suffers from a serious drawback, i.e., the joint-hypothesis problem \citep{CLM1997,Fama1991}. For this reason, I rely on another concept which I call ``market sensitivity.'' A financial market is said to be \textit{sensitive} to $\textbf{F}$ if and only if $\textbf{E}$ is $\P$-immersed in $\textbf{F}$. This is a rigorous definition of informational efficiency in terms of martingale theory. Put another way, in a sensitive market, the evolution of asset prices ``fully reflects'' or ``rapidly adjusts to'' the information flow $\textbf{F}$. In Section \ref{Sec.: Sensitivity} I show that the concept of market sensitivity is intimately connected to different notions of the Efficient-Market Hypothesis. Nevertheless, sensitivity does not require that the market is a fair game and thus, in contrast to the classic approach to market efficiency, it does not suffer from the joint-hypothesis problem (see Section \ref{Sec.: The Classic Approach to Market Efficiency}).

A financial market is said to be \textit{arbitrage free} if and only if there is no free lunch with vanishing risk (NFLVR) and no dominance (ND) with respect to the information flow $\textbf{F}$. Due to the 1\textsuperscript{st} Fundamental Theorem of Asset Pricing (FTAP), the NFLVR condition alone only guarantees that there exists an equivalent probability measure $\Q$ such that each discounted price process is a \emph{local} $\Q$-martingale with respect to $\textbf{F}$ \citep{DS1994}. \citet{JL2012} prove that, in every simple market with finite lifetime, the additional ND condition turns the discounted price processes into $\Q$-martingales with respect to $\textbf{F}$. Conversely, if a simple market with finite lifetime contains an equivalent martingale measure $\Q$ with respect to $\textbf{F}$, it must be arbitrage free. This result is referred to as the 3\textsuperscript{rd} FTAP \citep{Jarrow2012}. In this work, I extend the 3\textsuperscript{rd} FTAP to financial markets with infinite lifetime.

Modern approaches to the Efficient-Market Hypothesis focus on the absence of arbitrage \citep{JL2012,Ross2005}. In fact, \citet{JL2012} show that NFLVR and ND \emph{together} are necessary and sufficient for the existence of a pure exchange economy, with finite lifetime and a finite number of assets, where all subjects use the information flow $\textbf{F}$ for their investment-consumption plans and the discounted price processes form an Arrow-Radner market equilibrium. This demonstrates that every simple market, with finite lifetime and symmetric information, that is considered ``efficient'' must be at least arbitrage free or, equivalently, the discounted price processes must be martingales with respect to $\textbf{F}$ under any equivalent probability measure $\Q$. Both the absence of arbitrage opportunities and the ability of asset prices to ``fully reflect'' or ``rapidly adjust to'' the information flow $\textbf{F}$ are fundamental assumptions of neoclassical finance \citep{Ross2005}. These axioms turn out to be essential also for the theory presented in this work and so I use the following definition of market efficiency: A financial market is said to be \textit{efficient} if and only if it is sensitive to $\textbf{F}$ and contains a \textit{risk-neutral measure}, i.e., an equivalent martingale measure $\Q$ with respect to $\textbf{F}$.\footnote{As a consequence of the extended version of the 3\textsuperscript{rd} FTAP, which I present in this work, the discounted price processes are even assumed to be uniformly integrable martingales under $\Q\,$.}

The mathematical tools I use belong to martingale theory \citep{JS2003} and the key results stem from a discipline called ``enlargement of filtrations,'' developed by \citet{YJ1978,YJ1985}.\footnote{For a nice overview see \citet[][Ch.\ 2]{Jeanblanc2010}, which contains a comprehensive list of references on that topic.} This is a popular instrument in modern finance and has often been applied in the recent literature, especially in the area of credit risk and insider trading \citep{Amendinger1999,BR2002,EJY2000,Kohatsu-Higa2007}. The enlargement of filtration is typically done under some probability measure $\Q$ that is equivalent to $\P$. To the best of my knowledge, the question of market sensitivity, where we are mainly concerned with an enlargement under the \emph{physical} measure, has not yet been investigated in the literature.

Since the 1\textsuperscript{st}, 2\textsuperscript{nd}, and 3\textsuperscript{rd} FTAP \citep{DS1994,DS1998,HP1981,HP1983,Jarrow2012,JL2012} are essential in this methodological framework, they are briefly discussed in Section \ref{Sec.: The 3rd FTAP} and Section \ref{Sec.: Completeness}. Another essential branch of literature is related to the benchmark approach propagated by \citet{PH2006}. This is based on the growth-optimal portfolio (GOP), which has been a subject of heated discussions \citep{Christensen2005,MTZ2011}. In fact, the benchmark approach goes back to \citet{Long1990}, who has introduced the notion of numéraire portfolio (NP). In Section \ref{Sec.: The Growth-Optimal Portfolio}, I give a short overview of the benchmark approach and explain the connection between the GOP and the NP.

The GOP plays a fundamental role in modern finance \citep{KK2007,MTZ2011,PH2006}. If the market contains no unbounded profit with bounded risk (NUPBR), the GOP can be used as an NP. Unfortunately, this leads only to a Law of Minimal Price. The question of how to obtain a Law of One Price, in the strict sense mentioned at the beginning of this introduction, has not yet been investigated in the literature. Section \ref{Sec.: The Martingale Hypothesis} contains the main result of this work. This can be put in a nutshell as follows:
\begin{quote}
\textsl{Every complete and sensitive market contains a specific numéraire such that\/ $\Q=\P$.}
\end{quote}

\section{Preliminary Definitions and Assumptions}

Let $\bigl(\Omega,\textbf{F},\P\bigr)$ be a filtered probability space where the filtration $\textbf{F}=\{\mathcal{F}_t\}_{t\geq0}$ is right-continuous and complete. It is implicitly assumed that $\mathcal{F}_\infty$ forms the $\sigma$-algebra of the given probability space. Consider an asset universe $\mathcal{A}$ with a finite or infinite number of \emph{primary} assets. Let $\mathcal{S}_t$ be the set of asset prices in $\mathcal{A}$ at time $t\geq0\,$. More precisely, it is supposed that $\{\mathcal{S}_t\}_{t\geq0}$ is an $\textbf{F}$-adapted price process. Two assets are considered identical if and only if their price processes coincide almost surely. For notational convenience, I omit the subscript ``$i\in I$'' in every expression of the form ``$\{X_i\}_{i\in I}$'' if the index set $I$ is clear from the context.

The filtration $\textbf{F}$ can be viewed as a cumulative flow of information evolving through time. Since $\{\mathcal{S}_t\}$ is $\textbf{F}$-adapted, $\mathcal{F}_t$ contains at least the price history $\mathcal{E}_t$ at every time $t\geq0\,$.\footnote{The fact that $\{\mathcal{S}_t\}$ is $\textbf{F}$-adapted does not imply that each market participant has access to the information flow $\textbf{F}$.} More precisely, $\mathcal{E}_t$ denotes the $\sigma$-algebra generated by the price history in $\mathcal{A}$ at time $t$. It is supposed that $\mathcal{E}_0$ is trivial, i.e., it contains only the $\P$-null and $\P$-one elements of $\mathcal{F}_\infty\,$. The evolution of asset prices is represented by $\textbf{E}=\{\mathcal{E}_t\}$, i.e., the natural filtration of the price process $\{\mathcal{S}_t\}$. A filtration $\textbf{I}=\{\mathcal{I}_t\}$ is said to be a \textit{subfiltration} if and only if $\textbf{E}\subseteq\textbf{I}\subseteq\textbf{F}$, i.e., $\mathcal{E}_t\subseteq\mathcal{I}_t\subseteq\mathcal{F}_t$ for all $t\geq0\,$.

The notation ``$X\in\mathcal{I}$'' means that the random quantity $X$ is $\mathcal{I}$-measurable, where $\mathcal{I}$ is any sub-$\sigma$-algebra of $\mathcal{F}_\infty$. Attributes that are ascribed to random quantities or stochastic processes are meant to hold almost surely. For example, the equality ``$X=Y$'' for any two random vectors $X$ and $Y$ means that each component of $X$ equals the corresponding component of $Y$ almost surely. Any inequality of the form ``$X\leq Y$,'' ``$X\geq Y$,'' ``$X<Y$,'' or ``$X>Y$'' is to be understood in the same sense. If $\{X_t\}$ is an $\R^d$-valued stochastic process, $\{X_t\}\geq a$ means that $\{X_t\}$ is almost surely (uniformly) bounded from below by $a\in\R^d$. Moreover, two stochastic processes are considered identical if and only if they coincide (almost surely).

Now, choose an arbitrary asset as a numéraire and let $\{S_{0t}\}$ be its price process. Every finite subset of $\mathcal{A}$ that contains the chosen numéraire asset plus $N\in\N$ other assets is said to be a \textit{subuniverse}.\footnote{In this work, the symbol ``$\N$'' stands for the set of positive integers, i.e., $\N=\{1,2,\ldots\}$.} This is symbolized by $A\subseteq\mathcal{A}$ and $S_t=\big(S_{0t},S_{1t},\ldots,S_{Nt}\big)$ denotes the corresponding vector of asset prices for all $t\geq0\,$. It is assumed that $\{S_t\}$ is a positive $\textbf{F}$-adapted $\R^{N+1}$-valued semimartingale being right-continuous with left limits (\textit{càdlàg}).\footnote{It is not assumed that $\{S_t\}$ is bounded or locally bounded.} Also its left-continuous version, i.e., $\{S_{t_-}\}$ (with $t_-=0$ for $t=0$), is assumed to be positive.

The limit of $\{S_t\}$, i.e., $S_\infty$, exists and is finite. Moreover, it is assumed that $S_\infty>0\,$. This general approach enables us to analyze markets with infinite lifetime. Markets with finite lifetime, e.g., the Black-Scholes model, can be considered a special case. This is simply done by assuming that $\mathcal{F}_t=\mathcal{F}_T$ for all $t\geq T$, where $T\in\;]\,0,\infty\,[$ is any fixed lifetime. Discrete-time financial markets are obtained in the same way, just by assuming that the filtration $\textbf{F}$ is constant over the time intervals $[t_i,t_{i+1}[$ for $i=0,1,\ldots,n-1$, $0=t_0<t_1<\ldots<t_n=T$, and $n\in\N$.

For notational convenience, but without loss of generality, it is supposed that $S_{i0}=1$ for $i=0,1,\ldots,N$. I usually refer to the $\R^{N+1}$-valued process of \emph{discounted} asset prices, i.e., $\{P_t\}$ with $P_t=\bigl(1,S_{1t}/S_{0t},\ldots,S_{Nt}/S_{0t}\bigr)$ for all $t\geq0\,$.\footnote{From It\^{o}'s Lemma it follows that $\bigl\{S^{-1}_{0t}\bigr\}$ is a semimartingale and the product of two semimartingales is also a semimartingale. This means $\{P_t\}$ is an $\R^{N+1}$-valued semimartingale.} Since $\{S_t\}$ and $\{S_{t_-}\}$ are assumed to be positive, we also have that $\{P_t\},\{P_{t_-}\}>0$. If I say that any statement is true for all $\{P_t\}$, I mean that it is true for the discounted price process in each subuniverse $A\subseteq\mathcal{A}$. Similarly, a statement is true for all $\{S_t\}$ if and only if it is true for the nominal price process in every $A\subseteq\mathcal{A}$. All previous statements are supposed to be true for all $\{S_t\}$ and $\{P_t\}$, respectively.

Every $\textbf{F}$-predictable $\R^{N+1}$-valued stochastic process $\{H_t\}$ with $H_t=(H_{0t},H_{1t},\ldots,H_{Nt})$ that is integrable with respect to the discounted price process $\{P_t\}$ is said to be a \textit{trading strategy}. The discounted value of the strategy at every time $t\geq0$ is given by
\[
V_t = \sum_{i=0}^N H_{it}P_{it} = V_0 + \int_0^t H_s\,\d P_s\,,
\]
where $V_0=\sum_{i=0}^N H_{i0}P_{i0}$ is the discounted initial value and $\int_0^t H_s\,\d P_s$ represents the discounted gain of the strategy up to time $t\geq0$.\footnote{Two strategies are considered identical if and only if their (discounted) value processes coincide.} This means $V_t$ evolves from \emph{self-financing} transactions between time $0$ and $t$. The integral $\int_0^tH_s\,\d P_s$ is to be understood in the sense of \citet[][p.~207]{JS2003}, i.e., as a stochastic \emph{vector} integral.\footnote{For this reason, the requirements on $\{H_t\}$ that are mentioned by \citet{HP1981} are too strict \citep{JM1991}. See also Remark 1.3 in \citet{Biagini2010}.}

The strategy $\{H_t\}$ is called \textit{admissible} if and only if there exists a real number $a\geq0$ such that $\big\{\int_0^tH_s\,\d P_s\big\}\geq-a$.\footnote{According to \citet[][Definition 2.7]{DS1994}, the strategy $\{H_t\}$ is called ``$a$-admissible'' if and only if $\big\{\int_0^tH_s\,\d P_s\big\}\geq-a$ for a given number $a>0$ but just ``admissible'' if and only if $\big\{\int_0^tH_s\,\d P_s\big\}\geq-a$ for \emph{some} $a\geq0$.} The discounted initial value of $\{H_t\}$, i.e., $V_0$, need not be constant. If we add $a-V_0$ numéraire assets at $t=0$, we obtain the strategy $\big\{H^a_t\big\}$, which has a nonnegative discounted value process $\big\{V^a_t\big\}$ with $V^a_t=a+\int_0^tH^a_s\,\d P_s$ for all $t\geq0$. In the case $a>0$ we can divide $\big\{H^a_t\big\}$ by $a$ so as to obtain the strategy $\big\{H^a_t/a\big\}$ whose discounted value process $\big\{V^a_t/a\big\}$ starts at 1 and remains nonnegative. By choosing a sufficiently high number $a$, we can even guarantee that both $\big\{V^a_t/a\big\}$ and its left-continuous version $\big\{V^a_{t_-}/a\big\}$ are positive. Each admissible strategy that leads to a positive discounted value process starting at 1, such that the left-continuous version of the discounted value process is positive, too, is said to be \textit{normalized}.\footnote{A normalized strategy is always 1-admissible by construction. Moreover, each normalized strategy is still normalized after any change of numéraire.} A normalization just leads to an affine-linear transformation of the discounted value process of $\{H_t\}$, which enables us to switch easily between the different no-arbitrage conditions explained in Section \ref{Sec.: No-Arbitrage Conditions}. This general framework shall guarantee that the basic assumptions of the fundamental theorems of asset pricing and of the benchmark approach are satisfied \citep{DS1994,DS1998,HP1981,HP1983,Jarrow2012,KK2007}.

In this work, we are often concerned with an equivalent martingale measure (EMM), an equivalent local martingale measure (ELMM) or an equivalent uniformly integrable martingale measure (EUIMM). A probability measure $\Q$ is said to be an E(L)MM with respect to $\textbf{F}$ if and only if
\begin{enumerate}
\item[(i)] $\Q$ is equivalent to $\P$ on $\mathcal{F}_\infty$ and
\item[(ii)] every discounted price process $\{P_t\}$ is a (local) $\Q$-martingale with respect to $\textbf{F}$.
\end{enumerate}
The equivalence between $\Q$ and $\P$ on $\mathcal{F}_\infty$ is denoted by $\Q\sim\P$. Further, $\mathcal{M}^a_A(\textbf{F})$ \big($\mathcal{U}^a_A(\textbf{F})$\big) is the set of all probability measures $\Q\sim\P$ such that the discounted price process $\{P_t\}$ in the subuniverse $A\subseteq\mathcal{A}$ is a (uniformly integrable) $\Q$-martingale with respect to $\textbf{F}$. The superscript ``$a$'' shall indicate the chosen numéraire asset $a\in\mathcal{A}$. Analogously, $\mathcal{L}^a_A(\textbf{F})$ denotes the set of all probability measures that are equivalent to $\P$ on $\mathcal{F}_\infty$ such that $\{P_t\}$ is a \emph{local} $\Q$-martingale with respect to $\textbf{F}$. Moreover, whenever I drop the subscript $A$, I mean that the corresponding martingale property holds for all $\{P_t\}$ in the given asset universe.

A statement like ``$\Q\in\mathcal{M}^a(\textbf{E})$'' does not imply that $\Q$ is equivalent to $\P$ on the $\sigma$-algebra $\mathcal{F}_\infty$ and even if $\Q\sim\P$, $\{P_t\}$ is not necessarily a $\Q$-martingale with respect to $\textbf{F}$. Nevertheless, we always have that $\mathcal{U}^a(\textbf{F})\subseteq\mathcal{M}^a(\textbf{F})\subseteq\mathcal{L}^a(\textbf{F})$ and
\[
\mathcal{L}^a(\textbf{F})\subseteq\mathcal{L}^a(\textbf{E}),\footnote{Since $\Q$ is equivalent to $\P$ on $\mathcal{F}_\infty$, it is also equivalent to $\P$ on $\mathcal{E}_\infty$. Due to \citet[][Theorem 3.6]{FP2011}, every positive $\textbf{E}$-adapted local $\Q$-martingale with respect to $\textbf{F}$ is a local $\Q$-martingale with respect to $\textbf{E}$.} \quad \mathcal{M}^a(\textbf{F})\subseteq\mathcal{M}^a(\textbf{E}),\footnote{Since $\{P_t\}$ is $\textbf{E}$-adapted, it holds that $\E_\Q(P_T\,|\,\mathcal{E}_t)=\E_\Q\bigl(\E_\Q(P_T\,|\,\mathcal{F}_t)\,|\,\mathcal{E}_t\bigr)=\E_\Q(P_t\,|\,\mathcal{E}_t) = P_t$ for all $0\leq t\leq T<\infty\,$.} \quad\text{and}\quad
\mathcal{U}^a(\textbf{F})\subseteq\mathcal{U}^a(\textbf{E}).\footnote{This is simply because uniform integrability does not depend on the chosen filtration.}
\]

Every probability measure $\Q\sim\P$ is associated with a unique Radon-Nikodym (derivative or density) process (RNP) $\{\Lambda_t\}$, i.e., a positive uniformly integrable $\P$-martingale with respect to $\textbf{F}$ with $\Lambda_\infty>0$. Although $\mathcal{F}_0$ need not be trivial, we can assume without loss of generality that $\Lambda_0=1$ (see Section \ref{Sec.: Radon-Nikodym Derivatives}). Each stochastic process $\{\Lambda_t\}$ that satisfies the aforementioned properties is said to be a (local) \emph{discount-factor process} (DFP) if and only if $\{\Lambda_tP_t\}$ is a (local) $\P$-martingale with respect to $\textbf{F}$ for every discounted price process $\{P_t\}$. Whenever the lifetime of the financial market is finite, the uniform-integrability assumption about $\{\Lambda_t\}$ can be dropped and it is clear that every DFP is a local DFP but not vice versa.\footnote{Each local DFP is a so-called \textit{local martingale deflator} (see Proposition \ref{Pr.: Local martingale deflator}).} Every (local) DFP $\{\Lambda_t\}$ has an associated probability measure $\Q\sim\P$ which is defined by
\[
\Q(F) = \int_F \Lambda_\infty\,\d\P\,,\qquad\forall~F\in\mathcal{F}_\infty\,.
\]
I say that $\{\Lambda_t\}$ is an $\textbf{F}$-RNP or a (local) $\textbf{F}$-DFP, respectively, to emphasize the underlying filtration $\textbf{F}$. Finally, each ratio $\Lambda_{t,T}=\Lambda_T/\Lambda_t$ ($0\leq t\leq T<\infty$) is said to be a \textit{discount factor} and I write $\Lambda_{t,\infty}=\Lambda_\infty/\Lambda_t$ for all $t\geq0\,$.

\begin{figure}
\begin{center}
\includegraphics[scale=.3]{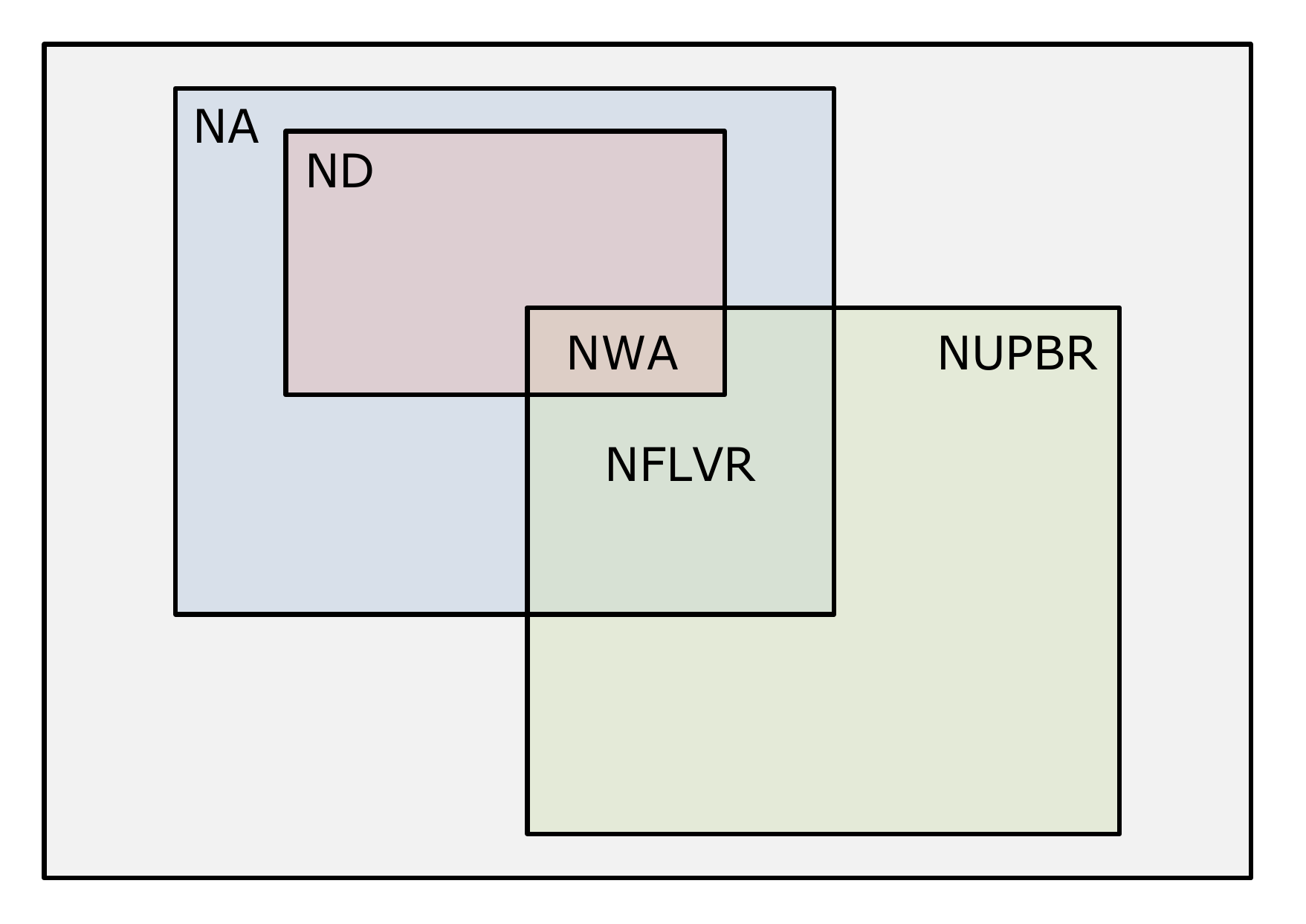}\\[-3ex]
\caption{Relationship between the several no-arbitrage conditions.\label{Fig.: No arbitrage}}
\end{center}
\end{figure}

In the following, I refer to several no-arbitrage conditions. Most of them are frequently applied in financial mathematics. Only the ND condition is not widespread in the literature. This no-arbitrage condition has been introduced by \citet{Merton1973} and can be found, e.g., in \citet{Jarrow2012} as well as \citet{JL2012}. All other no-arbitrage conditions are well-established. See for example \citet{KK2007} for a nice overview or consult Section \ref{Sec.: No-Arbitrage Conditions}.

A dominant strategy, a free lunch with vanishing risk, and an unbounded profit with bounded risk can be seen as weak arbitrage opportunities. I say that there is \textit{no weak arbitrage} (NWA) if and only if there is ND and NFLVR or, equivalently, ND and NUPBR, i.e.,
\[
\text{NWA}~:\Longleftrightarrow~\text{ND}~\wedge~\text{NFLVR}~\Longleftrightarrow~\text{ND}~\wedge~\text{NUPBR}\,.
\]
The relationship between the several no-arbitrage conditions is illustrated in Figure \ref{Fig.: No arbitrage}.

If the information flow $\textbf{F}$ does not allow for a weak arbitrage in the given subuniverse $A$, I say that $A$ is arbitrage free and write $\operatorname{NWA}^a_A(\textbf{F})$. The statements $\operatorname{NFLVR}^{\,a}_A(\textbf{F})$ and $\operatorname{NUPBR}^{\,a}_A(\textbf{F})$ shall be understood in the same sense. Moreover, the entire market or, equivalently, the asset universe $\mathcal{A}$, is said to be arbitrage free if and only if $\operatorname{NWA}^a_A(\textbf{F})$ for all $A\subseteq\mathcal{A}$. The distinction between $A$ and $\mathcal{A}$ is crucial if the market is complex.

\section{The Third Fundamental Theorem of Asset Pricing}\label{Sec.: The 3rd FTAP}

The 1\textsuperscript{st} FTAP for unbounded price processes \citep{DS1998} states that $\operatorname{NFLVR}^{\,a}_A(\textbf{F})$ if and only if $\{P_t\}$ is a $\Q$-$\sigma$-martingale with respect to $\textbf{F}$, where $\Q$ is equivalent to $\P$.\footnote{The stochastic process $\{Y_t\}$ is said to be a $\Q$-$\sigma$-martingale with respect to $\textbf{F}$ if and only if $Y_t=Y_0+\int_0^tH_s\,\d X_s$ for all $t\geq0$. Here $\{H_t\}$ is an $\{X_t\}$-integrable $\textbf{F}$-predictable stochastic process and $\{X_t\}$ is a local $\Q$-martingale with respect to $\textbf{F}$ (see Proposition 2 (i) in \citet{Emery1980} and Theorem III.6.41 in \citet[][p.~217]{JS2003}).} Every local martingale is a $\sigma$-martingale and every $\sigma$-martingale that is bounded from below is a local martingale \citep[][p.~214, 216]{JS2003}. Since the discounted asset prices are positive, $\{P_t\}$ is a local martingale if and only if it is a $\sigma$-martingale. For this reason, it is not necessary to distinguish between the terms ``local martingale'' and ``$\sigma$-martingale'' in the present context. This means $\operatorname{NFLVR}^{\,a}_A(\textbf{F})$ if and only if $\{P_t\}$ is a local $\Q$-martingale with $\Q\sim\P$. Every positive local martingale is a supermartingale. Hence, $P_t\geq\E_\Q(P_T\,|\,\mathcal{F}_t)$ for all $0\leq t\leq T<\infty$ and so the 1\textsuperscript{st} FTAP provides only a lower bound for the discounted price process.

Now, suppose that the financial market has a fixed finite lifetime $T>0\,$. In this situation, the 3\textsuperscript{rd} FTAP \citep{Jarrow2012} strengthens the 1\textsuperscript{st} FTAP. It states that there is NWA with respect to $\{\mathcal{F}_t\}_{0\leq t\leq T}$ if and only if $\{P_t\}$ is a $\Q$-martingale with respect to $\{\mathcal{F}_t\}$ for any $\Q\sim\P$. Moreover, \citet[][Theorem 3.2]{JL2012} show that the existence of an EMM with respect to $\{\mathcal{F}_t\}$ is equivalent to the existence of a pure exchange economy, with finite lifetime $T>0$, where all subjects use the same information flow $\{\mathcal{F}_t\}$ and $\{P_t\}$ is a discounted Arrow-Radner equilibrium-price process with respect to $\{\mathcal{F}_t\}$.\footnote{This means (i) the investment-consumption plans of all subjects are optimal with respect to $\{\mathcal{F}_t\}$ and (ii) all (i.e., the security and the commodity) markets clear with $\{P_t\}$.} Hence, the absence of weak arbitrage opportunities seems to be an essential requirement---not only for risk-neutral valuation but also for the existence of any market equilibrium in a finite economy.\footnote{Under short-selling constraints, a market equilibrium at least implies the existence of a local martingale deflator $\{\Lambda_t\}$. This guarantees that $\{\Lambda_tP_t\}$ is a local $\P$-martingale with respect to $\{\mathcal{F}_t\}$ \citep[][Theorem 3.1]{JL2013}.} This result marks a cornerstone in the development of the Efficient-Market Hypothesis.

The following theorem extends the 3\textsuperscript{rd} FTAP to financial markets with infinite lifetime.

\begin{theo}[The 3\textsuperscript{rd} FTAP]\label{Th.: The 3rd FTAP}
Let $A\subseteq\mathcal{A}$ be any subuniverse and $a\in A$ some numéraire asset. Then $\mathcal{U}^a_A(\textbf{F})\neq\emptyset$ if and only if\/ $\operatorname{NWA}^a_A(\textbf{F})$.
\end{theo}

Proof: If $\operatorname{NWA}^a_A(\textbf{F})$ there cannot exist a free lunch with vanishing risk with respect to $\textbf{F}$ in the subuniverse $A$ and thus we can apply Theorem 2.12 in \citet{DS1997} as well as Theorem 5.7 in \citet{DS1998}.\footnote{The admissibility condition given by \citet{DS1998} is always satisfied in this context and recall that we do not have to distinguish between $\sigma$-martingales and local martingales.} Since every asset in $A$ is $\textbf{F}$-maximal, it follows from Theorem 2.12 in \citet{DS1997} that the sum of all assets in $A$ is $\textbf{F}$-maximal, too.\footnote{\citet{JL2012} remark that the requirement that $\{P_t\}$ is locally bounded, which is given by \citet{DS1997}, in fact is superfluous.} Theorem 5.7 in \citet{DS1998} implies that there exists an ELMM $\Q$ with respect to $\textbf{F}$ such that the sum of all discounted asset prices in $A$ is a uniformly integrable $\Q$-martingale with respect to $\textbf{F}$. Hence, the discounted price process in $A$ is a positive local $\Q$-martingale bounded above by a uniformly integrable $\Q$-martingale and so it is also a uniformly integrable $\Q$-martingale with $\Q\sim\P$, i.e., $\mathcal{U}^a_A(\textbf{F})\neq\emptyset$. Conversely, if there exists a measure $\Q\sim\P$ such that the discounted price process of $A$ is a uniformly integrable $\Q$-martingale with respect to $\textbf{F}$, Theorem 5.7 in \citet{DS1998} implies that each asset in $A$ is $\textbf{F}$-maximal, whereas the 1\textsuperscript{st} FTAP guarantees that there is NFLVR with respect to $\textbf{F}$ in $A$. Hence, we have that $\operatorname{NWA}^a_A(\textbf{F})$.\QED

The uniform integrability of $\{P_t\}$ is an essential requirement. It leads to a financial market that is consistent in the following sense.

\begin{theo}[Change of numéraire]\label{Th.: Change of numéraire}
Let $a\in\mathcal{A}$ be some numéraire asset. If $\mathcal{U}^a(\textbf{F})\neq\emptyset$ then $\mathcal{U}^b(\textbf{F})\neq\emptyset$ for every other numéraire asset $b\in\mathcal{A}$.
\end{theo}

Proof: Let $\big\{S^a_t\big\}$ be the price process of the numéraire asset $a$ and $\big\{S^b_t\big\}$ the price process of any numéraire asset $b\neq a$. Further, consider an EUIMM $\Q\in\mathcal{U}^a(\textbf{F})$. Then $\big\{S^b_t/S^a_t\big\}$ is a positive uniformly integrable $\Q$-martingale with respect to $\textbf{F}$ with $S^b_0/S^a_0=1$ and $S^b_\infty/S^a_\infty>0$. Hence, we obtain the EMM $\widetilde \Q=\int \Gamma_\infty\,\d\Q\in\mathcal{M}^b(\textbf{F})$ with $\Gamma_t=S^b_t/S^a_t$ for all $t\geq0$. Since
\[
\frac{S_t}{S^b_t} = \E_\Q\left(\frac{\Gamma_\infty}{\Gamma_t}\,\frac{S_\infty}{S^b_\infty}\,|\,\mathcal{F}_t\right) = \E_{\widetilde\Q}\left(\frac{S_\infty}{S^b_\infty}\,|\,\mathcal{F}_t\right)
\]
for all $t\geq0$ and $\{S_t\}$, each $\widetilde\Q$-martingale $\big\{S_t/S^b_t\big\}$ is closed by $S_\infty/S^b_\infty$. This means $\widetilde\Q\in\mathcal{U}^b(\textbf{F})$, i.e., $\mathcal{U}^b(\textbf{F})\neq\emptyset$.\QED

The previous theorems justify the following definition.

\begin{defn}[Risk-neutral measure] Let $a\in\mathcal{A}$ be some numéraire asset. A probability measure $\Q$ is said to be a \emph{risk-neutral measure} if and only if\/ $\Q\in\mathcal{U}^a(\textbf{F})\,$.
\end{defn}

The existence of a risk-neutral measure implies that the market is arbitrage free in the sense of Theorem \ref{Th.: The 3rd FTAP}, i.e., that there is NWA with respect to $\textbf{F}$. Nonetheless, \citet{Herdegen2014} points out that most no-arbitrage conditions essentially depend on the choice of the numéraire asset. For this reason, he refrains from using a numéraire asset and suggests a numéraire-independent modeling framework for financial markets. Theorem \ref{Th.: Change of numéraire} at least guarantees that the existence of a risk-neutral measure is invariant under a change of numéraire. By contrast, $\mathcal{L}^a(\textbf{F})\neq\emptyset$ only guarantees that there is NFLVR with respect to $\textbf{F}$, but a change of numéraire can destroy the local martingale property \citep{DS1995b}.

The following theorem provides an equivalent representation of the discounted price process $\{P_t\}$ in terms of the real-world measure $\P$ instead of the risk-neutral measure $\Q$.

\begin{theo}[Representation Theorem]\label{Th.: Representation Theorem}
Let $A\subseteq\mathcal{A}$ be any subuniverse and $a\in A$ some numéraire asset. Then $\operatorname{NWA}^a_A(\textbf{F})$ if and only if\/ there exists an $\textbf{F}$-DFP $\{\Lambda_t\}$ such that $\{\Lambda_tP_t\}$ is a uniformly integrable\/ $\P$-martingale with respect to $\textbf{F}$.
\end{theo}

Proof: I start with the ``only if'' part. According to Theorem \ref{Th.: The 3rd FTAP}, $\operatorname{NWA}^a_A(\textbf{F})$ implies that $\mathcal{U}^a_A(\textbf{F})\neq\emptyset$. Consider some risk-neutral measure $\Q\in\mathcal{U}^a_A(\textbf{F})$ and let $\{\Lambda_t\}$ be the associated $\textbf{F}$-RNP. From Lemma \ref{Lem.: Change of martingale} we know that $\{\Lambda_tP_t\}$ is a $\P$-martingale with respect to $\textbf{F}$. Moreover, from Lemma \ref{Lem.: Change of measure} we conclude that
\[
P_t = \E_\Q(P_\infty\,|\,\mathcal{F}_t) = \E_\P\left(\frac{\Lambda_\infty}{\Lambda_t}\,P_\infty\,|\,\mathcal{F}_t\right)
\]
and thus $\Lambda_tP_t=\E_\P(\Lambda_\infty P_\infty\,|\,\mathcal{F}_t)$ for all $t\geq0\,$. Hence, the $\P$-martingale $\{\Lambda_tP_t\}$ is closed by $\Lambda_\infty P_\infty$ and thus uniformly integrable. Lemma \ref{Lem.: RNP <-> DFP} guarantees that $\{\Lambda_t\}$ is an $\textbf{F}$-DFP. For the ``if'' part consider the $\textbf{F}$-DFP $\{\Lambda_t\}$ and let $\Q\in\mathcal{M}^a(\textbf{F})$ be the associated EMM. Since $\{\Lambda_tP_t\}$ is uniformly integrable, we have that $\Lambda_tP_t=\E_\P(\Lambda_\infty P_\infty\,|\,\mathcal{F}_t)$ and with Lemma \ref{Lem.: Change of measure} we obtain
\[
P_t = \E_\P\left(\frac{\Lambda_\infty}{\Lambda_t}\,P_\infty\,|\,\mathcal{F}_t\right) = \E_\Q(P_\infty\,|\,\mathcal{F}_t)
\]
for all $t\geq0$. This means the $\Q$-martingale $\{P_t\}$ is closed by $P_\infty$ and thus it is uniformly integrable. We conclude that $\Q\in\mathcal{U}^a(\textbf{F})$ and from Theorem \ref{Th.: The 3rd FTAP} it follows that $\operatorname{NWA}^a_A(\textbf{F})$.\QED

So far, we have established the basic conditions for risk-neutral valuation, but some important issues are still missing on the agenda (see also p.~\pageref{T.: Problems (verbal)}):\label{T.: Problems (formal)}
\begin{enumerate}
\item[(i)] In real life, we do not know the set of risk-neutral measures, i.e., $\mathcal{U}^a(\textbf{F})$. In fact, this set might be considerably smaller than $\mathcal{U}^a(\textbf{E})$.
\item[(ii)] In general, $\mathcal{U}^a(\textbf{F})$ contains a multitude of risk-neutral measures and so the fair value of a contingent claim might not be unique, even if $\mathcal{U}^a(\textbf{F})$ was known.
\item[(iii)] Moreover, even if $\mathcal{U}^a(\textbf{F})$ is a singleton, it is practically impossible to derive the risk-neutral measure without making additional assumptions on the (discounted) price processes.
\end{enumerate}

Theorem \ref{Th.: Representation Theorem} is merely a re-formulation of Theorem \ref{Th.: The 3rd FTAP}. For this reason, the aforementioned obstacles cannot be cleared by the Representation Theorem: In general, (i) the DFP $\{\Lambda_t\}$ is not $\textbf{E}$-adapted, (ii) $\{\Lambda_t\}$ is not unique, and (iii) it is not a priori clear how to calculate $\{\Lambda_t\}$. In the following, I present the economic conditions under which $\{\Lambda_t\}$ turns out to be a unique and well-defined $\textbf{E}$-adapted stochastic process so that the aforementioned problems evaporate.

\section{Market Completeness and Sensitivity}

\subsection{Completeness}\label{Sec.: Completeness}

Consider a simple financial market with finite lifetime $T>0$ and choose any asset $a\in\mathcal{A}$ as a numéraire. \citet{HP1981} call every positive random variable $C\in\mathcal{E}_T$ a \textit{contingent claim}. They suppose that $\mathcal{M}^a(\textbf{E})\neq\emptyset$ and fix any $\Q\in\mathcal{M}^a(\textbf{E})$. Now, according to \citet{HP1981}, the financial market is \textit{complete} if and only if for every contingent claim $C$ with $\E_\Q(C/S^a_T)<\infty\,$, there exists an $\textbf{E}$-predictable strategy $\{H_t\}$ whose discounted value process $\{V_t\}$ is a $\Q$-martingale with respect to $\textbf{E}$ such that $V_T=C/S^a_T\,$. This implies that $\{V_t\}$ is positive. Moreover, by the Predictable Stopping Theorem \citep[][Lemma I.2.27]{JS2003}, also the left-continuous version of $\{V_t\}$ is positive. Since the $\sigma$-algebra $\mathcal{E}_0$ is assumed to be trivial, $V_0$ is constant and so the chosen strategy is admissible.

The economic idea behind the definition of market completeness can be explained like this: The goal is to replicate a contingent claim $C$ by an admissible $\textbf{E}$-predictable strategy $\{H_t\}$ as favorable as possible. Theorem 2.9 in \citet{DS1994} implies that $\{V_t\}$ must be a $\Q$-supermartingale with respect to $\textbf{E}$. This means we have that $V_t\geq\E_\Q\big(C/S^a_T\,|\,\mathcal{E}_t\big)$ for all $0\leq t\leq T$. Hence, in a complete financial market, we achieve the best possible replicating strategy if and only if the resulting discounted value process is a $\Q$-martingale with respect to $\textbf{E}$. We conclude that the \textit{fair value} of $C$ (expressed in units of the basic currency) amounts to $S^a_t\E_\Q\big(C/S^a_T\,|\,\mathcal{E}_t\big)$ at every time $0\leq t\leq T$. It is worth emphasizing that calculating the fair value of a contingent claim $C$ makes no sense if the market already contains an asset with discounted price process $\{\Pi_t\}$ such that $\Pi_T=C/S^a_T$. In this case, we can already observe the (nominal) price of the contingent claim at every time $0\leq t\leq T$ and, since we have that $\mathcal{M}^a(\textbf{E})\neq\emptyset$, this can be considered a fair value of $C$.

The 2\textsuperscript{nd} FTAP \citep{HP1983} states that a market is complete if and only if $\Q$ is the \emph{unique} EMM with respect to $\textbf{E}$. Moreover, it is complete if and only if $\{P_t\}$ satisfies the predictable-representation property. This means every $\Q$-martingale $\{X_t\}$ with respect to $\textbf{E}$ can be represented by $X_t=X_0+\int_0^t H_s\,\d P_s$ for all $0\leq t\leq T$, where $\{H_t\}$ is an $\textbf{E}$-predictable (not necessarily admissible) trading strategy. Unfortunately, in the continuous-time framework, only a small number of market models satisfy the desired predictable-representation property.

It is not meaningful to expand the concept of market completeness from $\textbf{E}$ to $\textbf{F}$ simply by substituting $\textbf{E}$ with $\textbf{F}$. In this case, we could only guarantee that every contingent claim $C\in\mathcal{E}_T$ is replicable by an $\textbf{F}$-predictable trading strategy, but this is not necessarily $\textbf{E}$-predictable. This means a market that is complete with respect to $\textbf{F}$ might be incomplete with respect to $\textbf{E}$. Put another way, if we substitute $\textbf{E}$ with $\textbf{F}$, market completeness would lack the so-called \emph{subset property} \citep{Latham1986}.\footnote{Here, I use the term ``subset property'' in a broad sense, albeit \citet{Latham1986} focuses on market efficiency.} The subset property is a natural requirement and turns out to be crucial when switching between the filtrations $\textbf{E}$ and $\textbf{F}$, which is frequently done in this work. Moreover, by substituting $\textbf{E}$ with $\textbf{F}$ we would allow $C$ to be an $\mathcal{F}_T$-measurable payoff, but in most practical situations it is sufficient and, for technical reasons, even necessary to assume that $C$ is determined only by the price history at time $T$, i.e., $C\in\mathcal{E}_T\,$.\footnote{For example, the Black-Scholes model requires that $\textbf{F}$ coincides with the natural filtration $\textbf{E}$ \citep{HP1981,JM1991}.} Interestingly, \citet[][p.~220]{HP1981} mention that they consider only the natural filtration $\textbf{E}$, whereas in \citet{HP1983} this essential point has been dropped.

If there exists a risk-neutral measure $\Q$ with respect to $\textbf{F}$ it is not sufficient to require that the discounted value process $\{V_t\}$ of the $\textbf{E}$-predictable strategy $\{H_t\}$ is a $\Q$-martingale with respect to $\textbf{E}$. More precisely, when replicating $C$ it should be possible to produce a $\Q$-martingale with respect to $\textbf{F}$. In this case, the replicating strategy $\{H_t\}$ is also fair with respect to $\textbf{F}$ although it is only $\textbf{E}$-predictable. This can be seen as follows: Suppose that we would allow $\{H_t\}$ to be an admissible $\textbf{F}$-predictable and not only $\textbf{E}$-predictable strategy. From Theorem 2.9 in \citet{DS1994} we conclude that the discounted value process of $\{H_t\}$ is a $\Q$-supermartingale with respect to $\textbf{F}$, i.e., $V_t \geq \E_\Q\big(C/S^a_T\,|\,\mathcal{F}_t\big)$ for all $0\leq t\leq T$, where $\E_\Q\big(C/S^a_T\,|\,\mathcal{F}_t\big)$ is the discounted value of the most favorable $\textbf{E}$-predictable replicating strategy at time $t$. This means we cannot find a better result by allowing the replicating strategy $\{H_t\}$ to be $\textbf{F}$-predictable.

The following definition of market completeness is based on the aforementioned arguments and is less restrictive than the original one. It allows for complex financial markets with infinite lifetime and an arbitrary filtration $\textbf{F}\supseteq\textbf{E}$. In particular, it satisfies the desired subset property. Thus it can be considered a natural generalization of the definition of market completeness given by \citet{HP1981,HP1983}.

\begin{defn}[Complete market]\label{Def.: Complete market}
Let $a\in\mathcal{A}$ be some numéraire asset and suppose that $\mathcal{U}^a(\textbf{F})\neq\emptyset$. Fix any risk-neutral measure $\Q\in\mathcal{U}^a(\textbf{F})$. The financial market is said to be \emph{complete} if and only if for every contingent claim $C\in\mathcal{E}_\infty$ with $\E_\Q(C/S^a_\infty)<\infty\,$, there exists an $\textbf{E}$-predictable strategy $\{H_t\}$ such that $V_t=\E_\Q(C/S^a_\infty\,|\,\mathcal{F}_t)$ for all $t\geq0$, where $\{V_t\}$ is the discounted value process of\/ $\{H_t\}$.
\end{defn}

The requirement of a risk-neutral measure is motivated by Theorem \ref{Th.: The 3rd FTAP}. Definition \ref{Def.: Complete market} allows $\{H_t\}$ to be based on \emph{any} subuniverse $A\subseteq\mathcal{A}$ of the financial market and it is assumed that the contingent claim $C$ is $\mathcal{E}_\infty$-measurable. Moreover, the strategy $\{H_t\}$ must be $\textbf{E}$-predictable and thus its discounted initial value $V_0$ is constant. The discounted value process $\{V_t\}$ is a uniformly integrable $\Q$-martingale with respect to $\textbf{F}$ (and not only with respect to $\textbf{E}$). This implies that $\{V_t\}>0$ and $\{V_{t_-}\}>0$, i.e., $\{H_t\}$ is admissible. Moreover, it follows that $V_\infty=C/S^a_\infty$ and so the given strategy indeed replicates the contingent claim $C$.

The chosen definition of market completeness is relatively weak. Since it is only required that the contingent claim is $\mathcal{E}_\infty$-measurable, we need not assume that it is possible to assess the fair value of any exotic derivative based on events that go beyond the history of asset prices. Typical examples are weather derivatives or non-financial bets. Nonetheless, market completeness does not \emph{exclude} the possibility to replicate (some) exotic instruments. Moreover, for a complex and complete market it is neither necessary nor sufficient that any finite subset of the asset universe forms a complete market. This means in a complete financial market, with an infinite number of assets, the predictable-representation property need not be satisfied in any subuniverse $A\subseteq\mathcal{A}$. In particular, every subuniverse might contain a multitude of equivalent martingale measures. The most striking example of a complex market, which is complete but model independent, is a ``dense'' market, i.e., a financial market where each contingent claim $C\in\mathcal{E}_\infty$ can be attained by a single asset. Note that the properties required by Definition \ref{Def.: Complete market} are implicitly satisfied for every $\textbf{E}$-predictable buy-and-hold single-asset strategy.

An important consequence of Definition \ref{Def.: Complete market} is that, for calculating the fair value of a contingent claim $C$, we need only the information flow $\textbf{E}$ but not the broader information flow $\textbf{F}$. On the one hand, the replicating strategy $\{H_t\}$ is only $\textbf{E}$-predictable and, on the other hand, it holds that $V_t=\E_\Q\big(C/S^a_\infty\,|\,\mathcal{F}_t\big)=\E_\Q\big(C/S^a_\infty\,|\,\mathcal{E}_t\big)$ for all $t\geq0$. Hence, if the market is complete with respect to $\textbf{F}$, each information that goes beyond the evolution of asset prices, $\textbf{E}$, but does not exceed the general information flow $\textbf{F}$ can be neglected. This solves the first part of the fundamental problem mentioned at the beginning of the introduction. The second part of the problem is solved by the following theorem.

\begin{theo}[Uniqueness]\label{Th.: Uniqueness}
Let $a\in\mathcal{A}$ be some numéraire asset. If the financial market is complete, $\mathcal{U}^a(\textbf{F})$ is a singleton.
\end{theo}

Proof: Since $\mathcal{U}^a(\textbf{F})\subseteq\mathcal{U}^a(\textbf{E})$ and $\mathcal{U}^a(\textbf{F})\neq\emptyset$, it follows that $\mathcal{U}^a(\textbf{E})\neq\emptyset$. Let $\{\Lambda_t\}$ be the $\textbf{E}$-RNP associated with any $\Q\in\mathcal{U}^a(\textbf{E})$. The market is complete and so the contingent claim $S^a_\infty\Lambda^{-1}_\infty>0$ can be attained by an $\textbf{E}$-predictable trading strategy with discounted value process $\{V_t\}$. We have that $V_\infty=\Lambda^{-1}_\infty$ and thus
\[
V_t = \E_\Q\bigl(V_\infty\,|\,\mathcal{F}_t\bigr) = \E_\Q\bigl(V_\infty\,|\,\mathcal{E}_t\bigr) = \E_\P\bigl(\Lambda_{t,\infty}V_\infty\,|\,\mathcal{E}_t\bigr) = \E_\P\bigl(\Lambda_{t,\infty}\Lambda^{-1}_\infty\,|\,\mathcal{E}_t\bigr) = \Lambda^{-1}_t
\]
for all $t\geq0\,$. Now, suppose that there exist two probability measures $\Q_1,\Q_2\in\mathcal{U}^a(\textbf{E})$ and let $\{\Lambda_{1t}\}$ and $\{\Lambda_{2t}\}$ be the associated $\textbf{E}$-RNPs. Then $\bigl\{\Lambda^{-1}_{1t}\bigr\}$ and $\bigl\{\Lambda^{-1}_{2t}\bigr\}$ are the discounted value processes for the contingent claims $S^a_\infty\Lambda^{-1}_{1\infty}$ and $S^a_\infty\Lambda^{-1}_{2\infty}\,$. Define $Q_t=\Lambda^{-1}_{1t}/\Lambda^{-1}_{2t}$ for all $t\geq0\,$. We see that both $\{Q_t\}$ and $\big\{Q^{-1}_t\big\}$ are $\P$-martingales with $Q_0=Q^{-1}_0=1$ and thus $\E_\P\bigl(Q_t\bigr)=\E_\P\bigl(Q^{-1}_t\bigr)=1$ for all $t\geq0\,$. This means we have that
\[
\E_\P\left(\frac{1}{Q_t}\right) = \frac{1}{\E_\P\bigl(Q_t\bigr)}\,,\qquad\forall~t\geq0\,.
\]
Since the function $f\!:x\mapsto x^{-1}$ for all $x>0$ is strictly convex, Jensen's inequality implies that $Q_t=1$ and thus $\Lambda_{1t}=\Lambda_{2t}$ for all $t\geq0\,$. This means $\mathcal{U}^a(\textbf{E})$ must be a singleton and so $\mathcal{U}^a(\textbf{F})\subseteq\mathcal{U}^a(\textbf{E})$ is a singleton, too.\QED

Theorem \ref{Th.: Uniqueness} states that each complete financial market cannot have more than one risk-neutral measure. This result holds irrespective of whether the market contains a finite or infinite number of assets. Similar statements can be found, e.g., in \citet{JM1999,JJM1999} as well as \citet{Biagini2010}. Hence, in every complete financial market we are always able to find a unique representation of asset prices and fair values.

The following theorem guarantees that market completeness does not depend on the chosen numéraire asset.

\begin{theo}[Change of numéraire]
Let $a\in\mathcal{A}$ be some numéraire asset. If the market is complete with respect to $a$ it is also complete with respect to every other numéraire asset $b\in\mathcal{A}$.
\end{theo}

Proof: Let $\Q\in\mathcal{U}^a(\textbf{F})$ be the risk-neutral measure and choose any other numéraire asset $b\in\mathcal{A}$. According to the proof of Theorem \ref{Th.: Change of numéraire}, we have that $\widetilde \Q=\int \Gamma_\infty\,\d\Q\in\mathcal{U}^b(\textbf{F})$ with $\Gamma_t=S^b_t/S^a_t$ for all $t\geq0\,$. Consider any contingent claim $C\in\mathcal{E}_\infty$ with $\E_{\widetilde\Q}\big(C/S^b_\infty\big)<\infty$. It holds that
\[
\E_\Q\big(C/S^a_\infty\big) = \E_\Q\left(\Gamma_\infty\,\frac{C}{S^b_\infty}\right) = \E_{\widetilde\Q}\big(C/S^b_\infty\big) < \infty\,.
\]
This means the contingent claim $C$ can be attained by an $\textbf{E}$-predicable strategy $\{H_t\}$ with value process $\{V_t\}$---discounted by $a$---such that $V_t=\E_\Q\big(C/S^a_\infty\,|\,\mathcal{F}_t\big)$ for all $t\geq0$. Now, given the numéraire asset $b$, the \emph{same} strategy leads to the discounted value process $\big\{\tilde V_t\big\}$ with
\[
\tilde V_t = \frac{S^a_t}{S^b_t} V_t = \E_\Q\left(\frac{\Gamma_\infty}{\Gamma_t}\frac{C}{S^b_\infty}\,|\,\mathcal{F}_t\right) = \E_{\widetilde\Q}\left(\frac{C}{S^b_\infty}\,|\,\mathcal{F}_t\right)
\]
for all $t\geq0$. We conclude that the market is complete with respect to $b\in\mathcal{A}$.\QED

The third part of the fundamental problem discussed on p.~\ref{T.: Problems (verbal)} and p.~\ref{T.: Problems (formal)} is still unsolved. This means I need to clarify the circumstances under which it is possible to represent asset prices and fair values in terms of $\P$. Put another way, we are waiting to see the (additional) condition that enables us to use the real-world measure as a \emph{risk-neutral} measure.

\subsection{Sensitivity}\label{Sec.: Sensitivity}

In the following, the time $t\geq0$ shall be understood as the ``present,'' every $s\geq0$ before $t$ is the ``past,'' whereas $T>t$ symbolizes the ``future,'' i.e., we have that $0\leq s\leq t<T<\infty$ unless otherwise stated. Let $X$ be some $\mathcal{E}_\infty$-measurable random vector. For example, $X$ could be a vector of asset prices, or any other function of asset prices, that will be manifested in the future. The complement of $\mathcal{E}_t$ relative to $\mathcal{F}_t$, i.e., $\sigma(\mathcal{F}_t\setminus\mathcal{E}_t)$, represents the information in $\mathcal{F}_t$ that goes beyond the price history $\mathcal{E}_t$. For example, if $\mathcal{F}_t$ is the set of public information then $\sigma(\mathcal{F}_t\setminus\mathcal{E}_t)$ denotes the subset of public information that does not belong to the price history at time $t$.

A natural requirement arising in financial econometrics is
\begin{equation}\label{Eq.: Irrelevance}
\P(X\leq x\,|\,\mathcal{F}_t) = \P(X\leq x\,|\,\mathcal{E}_t)
\end{equation}
for all $t\geq0\,$, $x\in\R^m$, $m\in\N$, and $\mathcal{E}_\infty$-measurable $m$-dimensional random vectors $X$. Eq.\ \ref{Eq.: Irrelevance} implies that the random vector $X$ is $\P$-independent of $\mathcal{F}_t$ \emph{conditional} on the price history $\mathcal{E}_t$. This means the conditional distribution of future asset prices might depend on the current history $\mathcal{E}_t$ of asset prices but not on any \emph{additional} information contained in $\mathcal{F}_t$. Under these circumstances, it is impossible to produce a better prediction of future asset prices (or functions thereof) by using some information in $\mathcal{F}_t$, provided the price history $\mathcal{E}_t$ has already been taken into account. More precisely, we have that
\[
\E_\P(X\,|\,\mathcal{F}_t) = \E_\P(X\,|\,\mathcal{E}_t)
\]
for all $t\geq0$ and $X\in\mathcal{E}_\infty$ with $\E_\P(|X|)<\infty\,$. Nevertheless, although it is superfluous to use any kind of information that exceeds $\mathcal{E}_t$ but is contained in $\mathcal{F}_t$, there might exist some information $\mathcal{G}_t$ \emph{beyond} $\mathcal{F}_t$ that could be useful.

Another desirable property is
\begin{equation}\label{Eq.: Rapid adjustment}
\P\bigl(Y_t\leq y\,|\,\mathcal{E}_\infty\bigr) = \P\bigl(Y_t\leq y\,|\,\mathcal{E}_t\bigr)
\end{equation}
for all $t\geq0\,$, $y\in\R^n$, $n\in\N$, and $\mathcal{F}_t$-measurable $n$-dimensional random vectors $Y_t\,$. For example, let $Y_t$ be a variable that indicates whether a stock company has committed a balance-sheet fraud up to time $t\geq0$ ($Y_t=1$) or not ($Y_t=0$). Since the choice of $\textbf{F}\supseteq\textbf{E}$ is arbitrary, we can suppose without loss of generality that $Y_t\in\mathcal{F}_t$ for all $t\geq0$. Consider an investor who takes only the current price history into account and is not aware of the fraud. It is assumed that the fraud will \emph{eventually} have an impact on the stock price. Hence, it would be ideal for the investor to know the future price evolution  \emph{today}, since on the basis of the future price movements, he or she would get a better assessment of the fraud probability. Unfortunately, in real life, $\mathcal{E}_\infty$ is unknown at time $t$. Nonetheless, Eq.~\ref{Eq.: Rapid adjustment} states that the investor can readily substitute $\mathcal{E}_\infty$ by $\mathcal{E}_t$. This means all information that would be useful for calculating the fraud probability, conditional on past and forthcoming price data, is already incorporated in the asset prices that can be observed \emph{now}. This paraphrases the widely accepted idea that asset prices ``rapidly adjust to'' new information \citep{FFJR1969}.

The following definition \citep[][p.~16]{Jeanblanc2010} is crucial for the subsequent analysis.

\begin{defn}[Immersion] Let $\Q$ be any probability measure. The filtration $\textbf{E}$ is said to be\/ $\Q$-\emph{immersed} in $\textbf{F}$ if and only if every square-integrable $\Q$-martingale with respect to $\textbf{E}$ is a square-integrable $\Q$-martingale with respect to $\textbf{F}$.
\end{defn}

The statement that ``$\textbf{E}$ is immersed in $\textbf{F}$'' (with respect to a probability measure $\Q$) is often referred to as the \textit{H-Hypothesis} \citep{BY1978}.

The following theorem provides different characterizations of the H-Hypothesis under the physical measure $\P$.

\begin{theo}[H-Hypothesis]\label{Th.: H-Hypothesis}
The following assertions are equivalent:
\begin{enumerate}
  \item[(i)] $\textbf{E}$ is\/ $\P$-immersed in $\textbf{F}$.
  \item[(ii)] It holds that $\P(X\leq x\,|\,\mathcal{F}_t) = \P(X\leq x\,|\,\mathcal{E}_t)$ for all $t\geq0\,$, $x\in\R^m$, $m\in\N$, and $m$-dimensional random vectors $X\in\mathcal{E}_\infty\,$.
  \item[(iii)] It holds that $\P\bigl(Y_t\leq y\,|\,\mathcal{E}_\infty\bigr)=\P\bigl(Y_t\leq y\,|\,\mathcal{E}_t\bigr)$ for all $t\geq0\,$, $y\in\R^n$, $n\in\N$, and $n$-dimensional random vectors $Y_t\in\mathcal{F}_t\,$.
  \item[(iv)] Every local\/ $\P$-martingale with respect to $\textbf{E}$ is a local\/ $\P$-martingale with respect to $\textbf{F}$.
\end{enumerate}
Moreover, if any one of the previous assertions is true it follows that
\[
\mathcal{E}_t=\mathcal{F}_t\cap\mathcal{E}_\infty\,,\qquad\forall~t\geq0\,.
\]
\end{theo}

Proof: Statements (i) to (iv) follow from Proposition 2.1.1 in \citet{Jeanblanc2010}. The last implication is part of Theorem 3 in \citet{BY1978}.\QED

Theorem \ref{Th.: H-Hypothesis} shows that the fundamental properties expressed by Eq.\ \ref{Eq.: Irrelevance} and Eq.\ \ref{Eq.: Rapid adjustment} are equivalent. This leads to the following definition.

\begin{defn}[Sensitive market] A financial market is said to be \emph{sensitive} if and only if any one of the equivalent assertions expressed by Theorem \ref{Th.: H-Hypothesis} is true. This is denoted by $\textbf{F}\rightsquigarrow\textbf{E}\,$.
\end{defn}

A financial market that is sensitive to $\textbf{F}$ is also sensitive to every subfiltration $\textbf{I}\,$. This means market sensitivity satisfies the subset property and $\textbf{F}\rightsquigarrow\textbf{E}$ does not exclude $\textbf{I}\rightsquigarrow\textbf{E}$ for any other filtration $\textbf{I}\supseteq\textbf{E}\,$. Moreover, it is trivial that $\textbf{E}\rightsquigarrow\textbf{E}\,$.

The following proposition provides a sufficient condition for market sensitivity.

\begin{prop}\label{Pr.: Sufficient condition}
Consider any probability measure $\Q\sim\P$ and let $\{\Lambda_t\}$ be the $\textbf{F}$-RNP associated with $\Q\,$. If $\textbf{E}$ is\/ $\Q$-immersed in $\textbf{F}$ and $\{\Lambda_t\}$ is $\textbf{E}$-adapted we have that $\textbf{F}\rightsquigarrow\textbf{E}\,$.
\end{prop}

Proof: This is a direct consequence of Proposition 2.1.4 in \citet{Jeanblanc2010}.\QED

There are many possibilities to define the meaning of informational efficiency in the sense that asset prices ``fully reflect'' some information flow $\textbf{F}$. For example, \citet{Dothan2008} states that,
\begin{quote}
``\textsl{The intuitive notion that prices fully reflect the information structure $F=(F_t)_{0\leq t\leq T}$ is then the requirement that the discounted price process $X_t$ be Markov.}''
\end{quote}
Unfortunately, the Markov assumption, i.e.,
\[
\P(X\leq x\,|\,\mathcal{E}_t) = \P(X\leq x\,|\,\mathcal{S}_t)\,,\qquad\forall~t\geq0\,,~x\in\R^m,~m\in\N,~X\in\mathcal{E}_\infty\,,
\]
essentially restricts the number of possible market models and it is well-known that this property is not satisfied in reality.\footnote{It is often supposed that $\P(X\leq x\,|\,\mathcal{F}_t)=\P(X\leq x\,|\,\mathcal{S}_t)$ ($\forall\,t\geq0,\,x\in\R^m,\,m\in\N,\,X\in\mathcal{E}_\infty$), which implies both market sensitivity \emph{and} the Markov property.}

The concept of market sensitivity is less restrictive, but it is still intimately connected to different notions of the Efficient-Market Hypothesis:

{\small
\begin{itemize}
  \item The relationship expressed by (\ref{Eq.: Irrelevance}) can be interpreted as a probabilistic definition of Fama's (\citeyear{Fama1970}) famous hypothesis that asset prices ``fully reflect'' $\mathcal{F}_t$ at every time $t\geq0\,$. For example, let $\mathcal{F}_t$ be the set of all private information at time $t$. If the market is strong-form efficient \citep{Fama1970} all private information, except for the price history $\mathcal{E}_t\,$, can be ignored because it is already ``incorporated'' in $\mathcal{E}_t\,$. Hence, if somebody aims at quantifying the conditional distribution of $X\in\mathcal{E}_\infty$, the weaker condition $\mathcal{E}_t$ is as good as the stronger condition $\mathcal{F}_t\,$, i.e., each private information beyond the price history is simply useless.
  \item The probability distribution of future asset prices generally depends on the underlying information. In a risky situation \citep{Knight1921}, the quality of each decision cannot become worse the more information is used.\footnote{This statement is no longer true under uncertainty \citep{Frahm2014}.} This means every market participant should gather as much information as possible.\footnote{This is true if the information costs are negligible \citep{GS1980}. Otherwise, each rational subject stops searching for information when the marginal cost approaches the marginal revenue \citep{Jensen1978}.} Consequently, the chosen market model must specify which kind of information is accessible by the economic subjects and used for their investment decisions. Suppose that their decisions are based only on the conditional distribution of future asset prices, i.e., other variables that will be manifested in the future do not matter. Eq.\ \ref{Eq.: Irrelevance} says that any information contained in $\mathcal{F}_t\,$, but being complementary to $\mathcal{E}_t\,$, would not alter the conditional price distribution and so this information can be simply ignored. More precisely, the economic subjects cannot improve their asset allocations by using some information in $\sigma(\mathcal{F}_t\setminus\mathcal{E}_t)$ provided they have already taken the current price history $\mathcal{E}_t$ into account. Hence, in a pure investment economy where Eq.\ \ref{Eq.: Irrelevance} is satisfied, the current asset prices would be unaffected by revealing $\mathcal{F}_t$ to all market participants. For example, if $\mathcal{F}_t$ is the set of private information, revealing some private information to the investors would not change their investment decisions and so the financial market is strong-form efficient in the sense of \citet{Latham1986} and \citet{Malkiel1992}.\footnote{Here, it is implicitly assumed that the subjects have already taken the current price history into account.}
  \item As already mentioned above, according to \citet{FFJR1969}, a financial market is considered efficient if it ``rapidly adjusts to'' new information. Eq.\ \ref{Eq.: Rapid adjustment} is the probabilistic counterpart of this statement and implies that every ``new information'' $F_t\in\mathcal{F}_t$ is instantaneously incorporated in the asset prices that can be observed at time $t$, i.e., \emph{now}, and not only at a later time $T>t$.
  \item \citet{Samuelson1965} conjectures that the market participants ``properly anticipate'' the future price evolution. He writes, ``If one could be sure that a price will rise, it would have already risen.'' Suppose that $\mathcal{E}_t\neq\mathcal{F}_t\cap\mathcal{E}_\infty$ for some $t\geq0\,$. Due to the last part of Theorem \ref{Th.: H-Hypothesis} it follows that the market is not sensitive. Hence, we have that $\mathcal{E}_t\subset\mathcal{F}_t\cap\mathcal{E}_\infty$ and so there exists an event $F_t\in\mathcal{F}_t\cap\mathcal{E}_\infty$ such that $F_t\not\in\mathcal{E}_t\,$. Since the event $F_t$ is also contained in $\mathcal{E}_\infty$ but exceeds $\mathcal{E}_t\,$, it leads to a situation where one can ``foresee'' to some degree the price evolution after time $t$. More precisely, the information $F_t$ reveals which sample paths are going to follow and which are not. This can be seen as a contradiction to Samuelson's doctrine. In the opposite case, i.e., if $\textbf{F}\rightsquigarrow\textbf{E}$ and thus $\mathcal{E}_t=\mathcal{F}_t\cap\mathcal{E}_\infty$ for all $t\geq0\,$, clairvoyance is impossible unless one has access to some information flow $\textbf{G}\supset\textbf{F}$ and the market is not sensitive to $\textbf{G}$.
\end{itemize}}\bigskip

The reason why the properties described by Theorem \ref{Th.: H-Hypothesis} characterize a ``sensitive'' market is best understood by examining a market that is \emph{not} sensitive. For this purpose, we have to take a closer look into the measure-theoretic framework. Let $E_t\in\mathcal{E}_t$ be the current history of asset prices and $E_T\in\mathcal{E}_T$ with $E_T\subset E_t$ the price history at some future point in time $T>t\,$. Suppose for the sake of simplicity that $\P(E_t)>\P(E_T)>0\,$. Consider a trader who operates on the basis of the information flow $\textbf{F}$ and let his or her investment decision at time $t$ be determined by the distribution of future asset prices conditional on $\mathcal{F}_t\,$. Since the market is not sensitive, we can assume that there exists some information $F_t\in\mathcal{F}_t$ with $F_t\subset E_t$ and $\P(F_t)>0$ such that $\P(E_T\,|\,F_t)\neq\P(E_T\,|\,E_t)$. In this case, the investment decision made by the trader, given the current history of asset prices, could depend on the realization of $\1_{F_t}\,$.\footnote{Here, $\1_{F_t}(\omega)=1$ if $\omega\in F_t$ and $\1_{F_t}(\omega)=0$ else ($\forall\,\omega\in\Omega$).} For example, the trader might want to buy some asset in case $\1_{F_t}=1$ but decides to sell the same asset if $\1_{F_t}=0\,$. By definition, the price history at time $t$ is $\mathcal{E}_t$-measurable, i.e., the past and current asset prices are \emph{constant} over the set $E_t\,$. Hence, the current asset prices are not \textit{sensitive} to $\1_{F_t}\,$, i.e., the trader is a \textit{price taker}---conditional on the current price history $E_t\,$. From an economic point of view, this is not desirable and characterizes a market where the asset prices do not ``fully reflect'' or ``rapidly adjust to'' the information flow $\textbf{F}$, although this flow of information could be useful also for other traders. Hence, perfect competition might enable a small investor to realize ``abnormal profits'' if he or she has access to information that is not already known to other investors. For example, this could be insider information.

We see that sensitivity is a highly desirable economic property. A market that is sensitive can immediately react to the news evolving with $\textbf{F}$, irrespective of whether those news are considered ``good'' or ``bad.'' This means in a sensitive market, the asset prices instantly adapt to the investment decisions that are based on the future price expectations of the market participants with respect to $\textbf{F}$. More precisely, each information $F_t\in\mathcal{F}_t$ that is considered \emph{useful} for assessing the physical distribution of future asset prices has an immediate impact on the supply and demand curves, which instantaneously affects the market quotes at time $t$. This does not mean that every subject who operates on the basis of $\textbf{F}$ makes the same investment decision. Market sensitivity does not even imply that the investment decisions are rational in any sense and pricing in a sensitive market need not be fair. For this reason, market sensitivity must not be confused with Fama's fair-game model \citep{Fama1970} or any other approach to market efficiency that requires the absence of ``economic profits'' \citep{Jensen1978}. Hence, the concept of market sensitivity does not suffer from the joint-hypothesis problem (see Section \ref{Sec.: The Classic Approach to Market Efficiency}).

Let $n\in\N$ be the number of market participants and suppose that each investor operates on the basis of some information flow $\{\mathcal{F}_{it}\}$ ($i=1,2,\ldots,n$). An ideal market is sensitive to the flow of private information, i.e., $\{\mathcal{G}_t\}$ with $\mathcal{G}_t=\sigma\big(\bigcup_{i=1}^n\mathcal{F}_{it}\big)$ for all $t\geq0\,$. If insider trading is prohibited and all insiders follow this rule, even an ideal market is not sensitive to the flow of insider information. Even if there exist a few insider traders, but the market is competitive, each insider is a price taker and so the market is still not sensitive to the flow of insider information. The bigger a group of investors acting on the same information flow, the greater its potential impact on the market prices. Hence, it can be assumed that financial markets are at least sensitive to the flow of public information, i.e., $\{\mathcal{I}_t\}$ with $\mathcal{I}_t=\bigcap_{i=1}^n\mathcal{F}_{it}$ for all $t\geq0\,$.

Market sensitivity \emph{per se} does not guarantee that the market is arbitrage free and in this specific sense ``efficient:'' If the market is sensitive but not arbitrage free, it is evident that all market participants will search in $\textbf{F}$ for arbitrage opportunities. No-arbitrage conditions only guarantee that the market is free of profits that would be realized by \emph{everyone}, irrespective of his or her own expectation, interest, and risk attitude. Nonetheless, if the market is arbitrage free but not sensitive, some market participants might still improve their positions by collecting data in addition to the current history of asset prices and re-allocating their capital. In either case, the market participants have an incentive to search for information that cannot be found just by investigating the history of asset prices. Only if the market is arbitrage free \emph{and} sensitive, it is impossible to ``make money out of nothing'' on the basis of $\textbf{F}$ \emph{and} the price evolution ``fully reflects'' or ``rapidly adjusts to'' the broader information flow $\textbf{F}$. The former is a fundamental assumption in financial mathematics, whereas the latter is a basic paradigm in finance theory. This justifies the following definition of market efficiency.

\begin{defn}[Efficient market]\label{Def.: Efficient market}
Let $a\in\mathcal{A}$ be some numéraire asset. The financial market is said to be \emph{efficient} if and only if $\mathcal{U}^a(\textbf{F})\neq\emptyset$ and $\textbf{F}\rightsquigarrow\textbf{E}\,$.
\end{defn}

Theorem \ref{Th.: Change of numéraire} guarantees that market efficiency does not depend on the chosen numéraire asset. Moreover, every complete and sensitive market is also efficient.

\section{The Growth-Optimal Portfolio}\label{Sec.: The Growth-Optimal Portfolio}

Fix a subuniverse $A\subseteq\mathcal{A}$ and choose any asset $a\in A$ as a numéraire. Further, let $\{H_t\}$ and $\{K_t\}$ be two \emph{normalized} $\textbf{F}$-predictable strategies whose discounted value processes are denoted by $\{V_t\}$ and $\{W_t\}$, respectively. Let $Q_t=V_t/W_t$ be the value of $V_t$ \textit{benchmarked} by $W_t$ at each time $t\geq0\,$. Since $Q_t=V_t/W_t=(S^a_tV_t)/(S^a_tW_t)$ for all $t\geq0$, it does not matter whether we express the values in units of the chosen numéraire asset or in units of the basic currency. This implies that the benchmarked value process $\{Q_t\}$ does not depend on the chosen numéraire asset at all.

The normalized strategy $\{K_t\}$ is said to be a \textit{numéraire portfolio} with respect to $\textbf{F}$ if and only if for every normalized $\textbf{F}$-predictable strategy $\{H_t\}$, the benchmarked value process $\{Q_t\}$ is a $\P$-supermartingale with respect to $\textbf{F}$, i.e., $\E_\P(Q_T\,|\,\mathcal{F}_t)\leq Q_t$ for all $0\leq t\leq T<\infty$. In particular, the stochastic process $\bigl\{W^{-1}_t\bigr\}$ is a positive $\P$-supermartingale. Doob's Martingale Convergence Theorem guarantees that $W^{-1}_\infty$ exists and is finite, i.e., $W^{-1}_\infty\in[0,\infty[\,$. Hence, the terminal value $W_\infty>0$ is well-defined, but we could have that $W_\infty=\infty$.\footnote{Since $\{Q_t\}$ is a nonnegative $\P$-supermartingale, it converges almost surely to some nonnegative random variable $Q_\infty\,$. This means also the discounted value process $\{V_t\}$ has a terminal value, i.e., $V_\infty=W_\infty Q_\infty\geq0\,$. In the unfavorable case $V_\infty=0\,$, the investor applies a so-called ``suicide strategy'' \citep{HP1981}.}

The strategy $\{K_t\}$ is said to be a \textit{growth-optimal portfolio} with respect to $\textbf{F}$ if and only if it maximizes the drift rate of $\{\log W_t\}$ with respect to $\textbf{F}$, i.e., the so-called \textit{growth rate} of $\{W_t\}$, for all $t\geq0$. Since $\log K'_tP_t=\log K'_tS_t-\log S^a_t$ for all $t\geq0$, every strategy is growth optimal with regard to the discounted price process $\{P_t\}$ if and only if it is growth optimal with regard to the nominal price process $\{S_t\}$. Thus growth optimality cannot be destroyed by moving from discounted to nominal asset prices and vice versa. Put another way, the choice of the numéraire does not matter for a GOP and the same holds for every NP.

A historical summary of the GOP is given by \citet{Christensen2005} and a rich collection of contributions related to the GOP can be found in \citet{MTZ2011}. \citet{KK2007} provide deep insights into the mathematical properties of the GOP and vividly explain its connection to the several no-arbitrage conditions discussed in this work.\footnote{See also \citet{HS2010} as well as \citet{IP2010} for similar results.} Theorem 3.15 in \citet{KK2007} describes a set of regularity conditions which guarantee that there exists one and only one GOP with respect to $\textbf{F}$. In this case, this is also an NP with respect to $\textbf{F}$. Conversely, if an NP with respect to $\textbf{F}$ exists, the regularity conditions are satisfied and the NP corresponds to the unique GOP with respect to $\textbf{F}$. Moreover, there exists an NP with \emph{finite} terminal value if and only if there is NUPBR \citep[][Theorem 4.12]{KK2007}. Throughout this section it is assumed that $\operatorname{NUPBR}^{\,a}_A(\textbf{F})$.

If the market is sensitive, every decision at time $t_-$ that is based on the conditional probability $\P(E\,|\,\mathcal{F}_{t_-})$ for any $E\in\mathcal{E}_\infty$ can be done as well on the basis of $\P(E\,|\,\mathcal{E}_{t_-})$. This can be seen as follows: Consider the two $\P$-martingales $\{M_t\}$ and $\{N_t\}$ with $M_t=\P(E\,|\,\mathcal{F}_t)$ and $N_t=\P(E\,|\,\mathcal{E}_t)$ for all $t\geq0$. Obviously, these martingales are identical if $\textbf{F}\rightsquigarrow\textbf{E}\,$. Now, the Predictable Stopping Theorem \citep[][Lemma I.2.27]{JS2003} implies that
\[
M_{t_-} = \E_\P(M_t\,|\,\mathcal{F}_{t_-}) = \P(E\,|\,\mathcal{F}_{t_-})\qquad\text{and}\qquad
N_{t_-} = \E_\P(N_t\,|\,\mathcal{E}_{t_-}) = \P(E\,|\,\mathcal{E}_{t_-})\,.
\]
Since $M_{t_-}=N_{t_-}$ for all $t\geq0$ (with $t_-=0$ for $t=0$), we have that
\[
\P(E\,|\,\mathcal{F}_{t_-}) = \P(E\,|\,\mathcal{E}_{t_-})\,,\qquad\forall~t\geq0,~E\in\mathcal{E}_\infty\,.
\]
In particular, the drift rates conditional on $\mathcal{F}_{t_-}$ and $\mathcal{E}_{t_-}$ coincide at every time $t\geq0$. Hence, if $\textbf{F}\rightsquigarrow\textbf{E}$ the GOP with respect to $\textbf{F}$ equals the GOP with respect to $\textbf{E}$.

As already mentioned, the GOP plays a fundamental role in modern finance. It serves as a benchmark portfolio \citep{Platen2006,Platen2009,PH2006}.\footnote{Under some additional assumptions, the GOP is a linear combination of the market portfolio and the money-market account \citep[][Ch.\ 11]{Platen2006,PH2006}. Typically, it is assumed that all market participants use the same information flow $\textbf{F}$ or at least that their expectations are rational.} There exists an important connection between the GOP and market sensitivity, which can be seen by the following theorem.

\begin{theo}[Benchmarked value process]\label{Th.: Benchmarked value process}
Suppose that\/ $\operatorname{NUPBR}^a_A(\textbf{F})$ and $\textbf{F}\rightsquigarrow\textbf{E}\,$. Let $\{V_t\}$ be the discounted value process of a normalized $\textbf{F}$-predictable strategy $\{H_t\}$ and $\{W_t\}$ the discounted value process of the GOP with respect to $\textbf{E}$. Further, let $\{Q_t\}$ with $Q_t=V_t/W_t$ for all $t\geq0$ be the benchmarked value process of $\{H_t\}$. Then
\begin{enumerate}
  \item[(i)] $\E_\P(Q_T\,|\,\mathcal{F}_t)\leq Q_t$ for all $0\leq t\leq T<\infty$ and
  \item[(ii)] for every $\sigma$-algebra $\mathcal{I}_t\subseteq\mathcal{F}_t$ and all\/ $0\leq t\leq T<\infty$ we have that
\[
\E_\P\left(\frac{Q_T}{Q_t}-1\,|\,\mathcal{I}_t\right) \leq 0\qquad\text{as well as}\qquad
\E_\P\left(\log\frac{Q_T}{Q_t}\,|\,\mathcal{I}_t\right) \leq 0.
\]
\end{enumerate}
\end{theo}

Proof: (i) Since the market is sensitive, $\{W_t\}$ represents the discounted value process of the NP with respect to $\textbf{F}$, which leads to the supermartingale property of $\{Q_t\}$. (ii) If we substitute $\mathcal{I}_t$ by $\mathcal{F}_t\,$, the first inequality is an immediate consequence of (i) and the second inequality follows from
\[
\E_\P\left(\log\frac{Q_T}{Q_t}\,|\,\mathcal{F}_t\right) \leq \log\E_\P\left(\frac{Q_T}{Q_t}\,|\,\mathcal{F}_t\right) \leq 0,\qquad\forall~0\leq t\leq T<\infty\,.
\]
The same inequalities with respect to $\mathcal{I}_t$ rather than $\mathcal{F}_t$ appear after applying the law of iterated expectations.\QED

Hence, if the market is sensitive, it is impossible to find a normalized $\textbf{F}$-predictable strategy whose \emph{benchmarked} value process leads to a positive expected (log-)return, conditional on some information set $\mathcal{I}_t\subseteq\mathcal{F}_t$ at any time $t\geq0$. The $\sigma$-algebra $\mathcal{I}_t$ need not contain $\mathcal{E}_t$. This means it can be any sub-$\sigma$-algebra of $\mathcal{E}_t$, e.g., the $\sigma$-algebra generated by a set of technical indicators or statistics based on the history of asset prices at time $t$. This allows us to apply simple hypothesis tests for market sensitivity and/or growth optimality. Here, I ignore the econometric implications of Theorem \ref{Th.: Benchmarked value process} and concentrate on aspects of financial mathematics.

Let $\{W_t\}$ be the discounted value process of the GOP with respect to $\textbf{F}$ and consider any contingent claim $C\in\mathcal{E}_T$ for a fixed time $T<\infty$ of maturity such that
\[
\E_\P\left(\frac{C/S^a_T}{W_T}\right) < \infty\,.
\]
Suppose that there is NFLVR with respect to $\textbf{F}$ and consider an admissible $\textbf{F}$-predictable strategy $\{H_t\}$ that leads to $C$, i.e., $V_T=C/S^a_T\,$. Since $C$ is positive, the discounted value process $\{V_t\}$ and its left-continuous version must be positive, too. This means $\{H_t/V_0\}$ is a normalized $\textbf{F}$-predictable strategy with discounted value process $\{V_t/V_0\}$. Hence, the benchmarked value process $\{(V_t/V_0)/W_t\}$ is a $\P$-supermartingale with respect to $\textbf{F}$ and thus
\[
\frac{V_t}{W_t} \geq \E_\P\left(\frac{C/S^a_T}{W_T}\,|\,\mathcal{F}_t\right)
\]
for all $0\leq t\leq T$. This means
\[
\left\{\E_\P\left(\frac{C/S^a_T}{W_T/W_t}\,|\,\mathcal{F}_t\right)\right\}_{0\leq t\leq T}
\]
forms a lower bound for the discounted value processes of \emph{all} admissible $\textbf{F}$-predictable strategies that lead to $C$. This means an investor who aims at replicating the payoff $C$, but has no more information than $\textbf{F}$, should try to choose an $\textbf{F}$-predictable strategy whose discounted value process attains the lower bound with respect to $\textbf{F}$. By contrast, if the investor has access to some broader information flow $\textbf{G}\supset\textbf{F}$, he or she might find a better strategy to obtain $C\,$. These arguments lead to the following definition \citep{Platen2009}.\footnote{\citet{Platen2009} only requires $\operatorname{NUPBR}^{\,a}_A(\textbf{F})$, but he explicitly assumes that $\{V_t\}$ is positive.}

\begin{defn}[Fair strategy] Suppose that $\operatorname{NFLVR}^{\,a}_A(\textbf{F})$ and let $\{W_t\}$ be the discounted value process of the GOP with respect to $\textbf{F}$. An admissible $\textbf{F}$-predictable strategy $\{H_t\}$ that leads to the terminal value $C\in\mathcal{E}_T$ for any fixed $T<\infty$ such that
\[
\E_\P\left(\frac{C/S^a_T}{W_T}\right) < \infty
\]
is said to be \emph{fair} with respect to $\textbf{F}$ if and only if
\[
V_t = \E_\P\left(\frac{C/S^a_T}{W_T/W_t}\,|\,\mathcal{F}_t\right),\qquad\forall~0\leq t\leq T\,.
\]
\end{defn}

From the previous arguments, we conclude that the discounted value process of a fair strategy as well as its left-continuous version is always positive.

In general, if some information flow $\textbf{F}$ is available to the investor, the GOP at time $t\in\big[0,T\big]$ should be calculated by $\mathcal{F}_{t_-}$ and not by $\mathcal{E}_{t_-}$, since otherwise he or she could overestimate the fair price of a contingent claim. By contrast, if the market is sensitive, using the information $\mathcal{E}_{t_-}$ is sufficient. This is the quintessence of the following theorem.

\begin{theo}[Fair strategy]\label{Th.: Fair strategy}
Suppose that $\operatorname{NFLVR}^a_A(\textbf{F})$ and $\textbf{F}\rightsquigarrow\textbf{E}\,$. If a strategy is fair with respect to $\textbf{E}$ it is fair with respect to $\textbf{F}$.
\end{theo}

Proof: Let $\{V_t\}$ be the discounted value process of a fair strategy $\{H_t\}$ with respect to $\textbf{E}$ that leads to $C\in\mathcal{E}_T$ and $\{W_t\}$ be the discounted value process of the GOP with respect to $\textbf{E}$, so that
\[
V_t = \E_\P\left(\frac{C/S^a_T}{W_T/W_t}\,|\,\mathcal{E}_t\right)
\]
for all $0\leq t\leq T$. Since the market is sensitive, we have that
\[
\E_\P\left(\frac{C/S^a_T}{W_T/W_t}\,|\,\mathcal{E}_t\right) = \E_\P\left(\frac{C/S^a_T}{W_T/W_t}\,|\,\mathcal{F}_t\right)
\]
for all $0\leq t\leq T$ and the GOP with respect to $\textbf{E}$ is also growth optimal with respect to $\textbf{F}\,$. Hence, $\{H_t\}$ is also fair with respect to $\textbf{F}$.\QED

This means an investor who aims at a positive payoff $C\in\mathcal{E}_T$ cannot gain anything by taking an information flow $\textbf{F}$ into account if the market is sensitive to $\textbf{F}$, provided he has already found a fair strategy with respect to $\textbf{E}$.\footnote{The proof of Theorem \ref{Th.: Fair strategy} reveals that the discounted value process of a fair strategy with respect to $\textbf{E}$ coincides with the discounted value process of a fair strategy with respect to $\textbf{F}$. This means the strategies are identical.}

\section{The Martingale Hypothesis}\label{Sec.: The Martingale Hypothesis}

The methodological framework that has been chosen in this work does not require a competitive market. In particular, it is not assumed that the market participants are price takers. This means any individual investment decision $H_{it_-}$ ($i=1,2,\ldots,n$) can have an influence on $\mathcal{S}_{t_-}$ and vice versa. Hence, the financial market can be highly illiquid. The price-taker assumption, i.e., the assumption that each order ``gets lost in the masses,'' does not adequately describe the pricing mechanism of financial markets. In fact, market sensitivity thrives on the fact that each investment decision \emph{has} an impact on the asset prices, whereas completeness guarantees that the market participants are able to tailor each financial instrument to their needs. Therefore, completeness and sensitivity mutually support each other and enables us to derive a simple real-world valuation formula. This is done in the subsequent analysis.

A method which comes very close to the target is the benchmark approach discussed in Section \ref{Sec.: The Growth-Optimal Portfolio}. Fix any subuniverse $A\subseteq\mathcal{A}$ with some numéraire asset $a\in A$. If there is $\operatorname{NUPBR}^{\,a}_A(\textbf{F})$, it must hold that
\begin{equation}\label{Eq.: Benchmark approach}
P_t \geq \E_\P\left(\frac{P_T}{W_T/W_t}\,|\,\mathcal{F}_t\right)
\end{equation}
for all $0\leq t\leq T<\infty$, where $\{W_t\}$ is the discounted value process of the GOP with respect to $\textbf{F}$. Unfortunately, the given result represents only a Law of Minimal Price \citep{Platen2009} but not a Law of One Price. The reason is twofold: (i) It provides only a lower bound for the discounted price process $\{P_t\}$ and (ii) even if (\ref{Eq.: Benchmark approach}) was an equality, the conditional expectation in general is not stable under a change of filtration. Another drawback is that for calculating the GOP with respect to $\textbf{F}$ it is not sufficient to take only asset prices into consideration. In general, it is necessary to search for data in $\textbf{F}$ that go beyond $\textbf{E}$.

In the following, I derive a Law of One Price under the assumption that the market is complete and sensitive. The basic idea is simple: I fix the physical measure $\P$ and search for a normalized $\textbf{E}$-predictable strategy $\star$ such that $\P\in\mathcal{U}^\star(\textbf{F})$. This means I treat the strategy $\star$ like an \emph{asset},\footnote{This is not to say that $\star$ belongs to the asset universe, $\mathcal{A}$, which contains only the \emph{primary} assets in the market.} which is possible only because $\star$ is determined by the evolution of asset prices. Hence, its value process is $\textbf{E}$-adapted, like every other price process. By contrast, risk-neutral valuation works the other way around: One fixes a numéraire asset $a\in\mathcal{A}$ and searches for some risk-neutral measure $\Q\in\mathcal{U}^a(\textbf{F})$ or at least for an EMM $\Q\in\mathcal{M}^a(\textbf{F})$.

The idea of fixing the physical measure and searching for an appropriate numéraire $\star$ such that $\P\in\mathcal{M}^\star(\textbf{F})$ or at least $\P\in\mathcal{L}^\star(\textbf{F})$ can be found in \citet{Becherer2001} and \citet{Long1990},\footnote{According to \citet{Becherer2001}, this approach even goes back to \citet[][p.~184]{Vasicek1977}.} but the results presented in this work differ in several aspects. The aforementioned authors (i) do not study conditions under which the discounted price processes turn out to be uniformly integrable $\P$-martingales with respect to $\textbf{F}$, (ii) do not distinguish between $\textbf{E}$ and $\textbf{F}$, and (iii) assume a financial market with finite lifetime, so that the essential requirement of uniform integrability becomes superfluous.

Most of the following results require a complete financial market. They are applicable both to simple and complex markets but, due to the arguments given in Section \ref{Sec.: Completeness}, it is tempting to think about a market with an infinite number of assets. This leads to a model-independent framework, i.e., although the market is assumed to be complete, it is not necessary to make any specific assumption about the (discounted) price processes.

\begin{prop}\label{Pr.: Main proposition}
Let $a\in\mathcal{A}$ be some numéraire asset and suppose that the financial market is complete. If the $\textbf{F}$-RNP associated with any\/ $\Q\in\mathcal{U}^a(\textbf{F})$ is $\textbf{E}$-adapted it follows that $\textbf{F}\rightsquigarrow\textbf{E}\,$.
\end{prop}

Proof: Consider any square-integrable $\Q$-martingale $\{X_t\}$ with respect to $\textbf{E}\,$. Hence, $\{X_t\}$ is uniformly $\Q$-integrable and converges to some limit $X_\infty\,$. Choose any real number $x>0$ and define
\[
X_{1\infty} := x+\max\{X_\infty\,,0\}>0\qquad\text{and}\qquad X_{2\infty} := x-\min\{X_\infty\,,0\}>0
\]
so that $X_\infty=X_{1\infty}-X_{2\infty}\,$. The market is complete and so the contingent claims $S^a_\infty X_{1\infty}>0$ and $S^a_\infty X_{2\infty}>0$ can be attained by two $\textbf{E}$-predictable trading strategies with discounted value processes $\{V_{1t}\}$ and $\{V_{2t}\}$. It follows that
\[
\E_\Q(V_{1\infty}\,|\,\mathcal{F}_t) = V_{1t} = \E_\Q(V_{1\infty}\,|\,\mathcal{E}_t)\qquad\text{and}\qquad
\E_\Q(V_{2\infty}\,|\,\mathcal{F}_t) = V_{2t} = \E_\Q(V_{2\infty}\,|\,\mathcal{E}_t)
\]
for all $t\geq0\,$. Hence, we obtain
\begin{align*}
X_t &= \E_\Q(X_\infty\,|\,\mathcal{E}_t) = \E_\Q(X_{1\infty}-X_{2\infty}\,|\,\mathcal{E}_t) = \E_\Q(X_{1\infty}\,|\,\mathcal{E}_t) - \E_\Q(X_{2\infty}\,|\,\mathcal{E}_t) \\
    &= \E_\Q(V_{1\infty}\,|\,\mathcal{E}_t) - \E_\Q(V_{2\infty}\,|\,\mathcal{E}_t) = \E_\Q(V_{1\infty}\,|\,\mathcal{F}_t) - \E_\Q(V_{2\infty}\,|\,\mathcal{F}_t) \\
    &= \E_\Q(X_{1\infty}\,|\,\mathcal{F}_t) - \E_\Q(X_{2\infty}\,|\,\mathcal{F}_t) = \E_\Q(X_{1\infty}-X_{2\infty}\,|\,\mathcal{F}_t) = \E_\Q(X_\infty\,|\,\mathcal{F}_t)
\end{align*}
for all $t\geq0$ and so $\{X_t\}$ is a (square-integrable) $\Q$-martingale with respect to $\textbf{F}\,$. This means $\textbf{E}$ is $\Q$-immersed in $\textbf{F}$. Now, Proposition \ref{Pr.: Sufficient condition} guarantees that $\textbf{F}\rightsquigarrow\textbf{E}\,$.\QED

The usual definition of the GOP can be applied to complex financial markets by allowing the investors to operate in any subuniverse of $\mathcal{A}$. A GOP based on a subuniverse $A\subseteq\mathcal{A}$ is simply said to be ``the GOP'' if and only if there is no GOP in any other subuniverse $B\subseteq\mathcal{A}$ that leads to a higher growth rate. The GOP remains growth optimal if prices and values are denominated in the basic currency. Moreover, as is shown in Section \ref{Sec.: The Growth-Optimal Portfolio}, every GOP with respect to $\textbf{E}$ is also growth optimal with respect to $\textbf{F}$ if the market is sensitive.

\begin{prop}\label{Pr.: Existence and uniqueness of the GOP}
Let $\star$ be any normalized $\textbf{E}$-predictable strategy and suppose that\/ $\P\in\mathcal{L}^\star(\textbf{F})$. Then $\star$ is the unique GOP with respect to $\textbf{F}$.
\end{prop}

Proof: Fix any subuniverse $A\subseteq\mathcal{A}$ containing the assets that are used by the strategy $\star$ and take $\star$ as a numéraire. Further, let $\{H_t\}$ be any normalized $\textbf{F}$-predictable strategy in $A$. From Theorem 2.9 in \citet{DS1994} it follows that the discounted value process of $\{H_t\}$ is a $\P$-supermartingale with respect to $\textbf{F}$. This means $\star$ is an NP in $A$ with respect to $\textbf{F}$. Theorem 3.15 in \citet{KK2007} implies that $\star$ is the unique GOP in $A$ with respect to $\textbf{F}$. The same holds for every other subuniverse that contains the assets of $\star$. Further, it is clear that any other subuniverse that does not contain all assets used by $\star$ cannot lead to a higher growth rate. This means $\star$ must be growth optimal with respect to $\textbf{F}$. By the same arguments, we may conclude that $\star$ is unique.\QED

\begin{theo}[Growth-optimal portfolio]\label{Th.: GOP}
Every complete and sensitive financial market contains a unique $\textbf{E}$-predictable GOP with respect to $\textbf{F}$.
\end{theo}

Proof: Consider some numéraire asset $a\in\mathcal{A}$. Let $\Q\in\mathcal{U}^a(\textbf{F})\subseteq\mathcal{U}^a(\textbf{E})$ be the unique risk-neutral measure and $\{\Lambda_t\}$ the $\textbf{E}$-RNP associated with $\Q$. From Lemma \ref{Lem.: RNP <-> DFP} we know that $\{\Lambda_t\}$ is an $\textbf{E}$-DFP, i.e., $\{\Lambda_tP_t\}$ is a $\P$-martingale with respect to $\textbf{E}$ for each discounted price process $\{P_t\}$. Since the market is sensitive, Theorem \ref{Th.: H-Hypothesis} implies that $\{\Lambda_tP_t\}$ is also a $\P$-martingale with respect to $\textbf{F}$ for each discounted price process $\{P_t\}$. Further, the market is complete and so there exists an $\textbf{E}$-predictable trading strategy $\{K_t\}$ with discounted value process $\{W_t\}$ such that $W_\infty=\Lambda^{-1}_\infty$ and
\[
W_t = \E_\Q(W_\infty\,|\,\mathcal{F}_t) = \E_\Q(W_\infty\,|\,\mathcal{E}_t) = \E_\P(\Lambda_{t,\infty}W_\infty\,|\,\mathcal{E}_t) = \E_\P(\Lambda_{t,\infty}\Lambda^{-1}_\infty\,|\,\mathcal{E}_t) = \Lambda^{-1}_t
\]
for all $t\geq0\,$. This means $\big\{P_t/W_t\big\}$ is a $\P$-martingale with respect to $\textbf{F}$ for each discounted price process $\{P_t\}$. Let $\{S^\star_t\}$ with $S^\star_t=S^a_tW_t$ for all $t\geq0$ be the nominal value process of $\{K_t\}$, so that each $\{S_t/S^\star_t\}$ is a $\P$-martingale with respect to $\textbf{F}$, i.e., $\P\in\mathcal{M}^\star(\textbf{F})\subseteq\mathcal{L}^\star(\textbf{F})$. Now, Proposition \ref{Pr.: Existence and uniqueness of the GOP} implies that $\{K_t\}$ is the unique GOP with respect to $\textbf{F}$.\QED

The following theorem is the main result of this work. It provides a simple characterization of market completeness and sensitivity. Moreover, it clarifies that under these ideal circumstances, the GOP with respect to $\textbf{F}$ is $\textbf{E}$-predictable and thus can be considered a ``benchmark asset.''

\begin{theo}[Martingale Hypothesis]\label{Th.: Martingale Hypothesis}
A complete financial market is sensitive if and only if there exists a normalized $\textbf{E}$-predictable strategy $\star$ such that\/ $\P\in\mathcal{U}^{\star}(\textbf{F})$. The strategy $\star$ corresponds to the unique GOP with respect to $\textbf{F}$.
\end{theo}

Proof: I start with the ``only if'' part. The proof of Theorem \ref{Th.: GOP} reveals that there exists an $\textbf{E}$-predictable GOP with respect to $\textbf{F}$ with nominal value process $\big\{S^\star_t\big\}$. Since the market is sensitive, we obtain
\[
\frac{S_t}{S^\star_t} = \E_\Q\left(\frac{W_\infty}{W_t}\,\frac{S_\infty}{S^\star_\infty}\,|\,\mathcal{F}_t\right) = \E_\Q\left(\frac{W_\infty}{W_t}\,\frac{S_\infty}{S^\star_\infty}\,|\,\mathcal{E}_t\right) = \E_\P\left(\frac{S_\infty}{S^\star_\infty}\,|\,\mathcal{E}_t\right) = \E_\P\left(\frac{S_\infty}{S^\star_\infty}\,|\,\mathcal{F}_t\right)
\]
for all $t\geq0$ and $\{S_t\}$, where $\Q\in\mathcal{U}^a(\textbf{F})$ represents the unique risk-neutral measure. This means each $\P$-martingale $\{S_t/S^\star_t\}$ is closed by $S_\infty/S^\star_\infty\,$, i.e., $\P\in\mathcal{U}^{\star}(\textbf{F})$. For the ``if'' part consider some numéraire asset $a\in\mathcal{A}$ and note that $\big\{W^{-1}_t\big\}$ with $W_t=S^\star_t/S^a_t$ for all $t\geq0$ is a positive uniformly integrable $\P$-martingale with respect to $\textbf{F}$ such that $W^{-1}_0=1$ and $W^{-1}_\infty>0\,$. This means $\big\{W^{-1}_t\big\}$ is an $\textbf{F}$-DFP with associated probability measure $\widetilde\Q=\int W^{-1}_\infty\,\d\P$ and we have that
\[
\frac{S_t}{S^a_t} = \E_\P\left(\frac{W^{-1}_\infty}{W^{-1}_t}\,\frac{S_\infty}{S^a_\infty}\,|\,\mathcal{F}_t\right) = \E_{\widetilde\Q}\left(\frac{S_\infty}{S^a_\infty}\,|\,\mathcal{F}_t\right)
\]
for all $t\geq0$ and $\{S_t\}$. Thus each $\widetilde\Q$-martingale $\{S_t/S^a_t\}$ is closed by $S_\infty/S^a_\infty\,$, i.e., $\widetilde\Q\in\mathcal{U}^a(\textbf{F})$. Since $\mathcal{U}^a(\textbf{F})$ is a singleton, we conclude that $\widetilde\Q=\Q\,$. This means $\big\{W^{-1}_t\big\}$ is the $\textbf{F}$-RNP associated with $\Q$ and, since $\big\{W^{-1}_t\big\}$ is $\textbf{E}$-adapted, Proposition \ref{Pr.: Main proposition} implies that $\textbf{F}\rightsquigarrow\textbf{E}$. Finally, from $\P\in\mathcal{U}^\star(\textbf{F})\subseteq\mathcal{L}^\star(\textbf{F})$ and Proposition \ref{Pr.: Existence and uniqueness of the GOP}, we conclude that $\star$ is the unique GOP with respect to $\textbf{F}$.\QED

Hence, every complete financial market is sensitive to the information flow $\textbf{F}$ if and only if the discounted price processes turn out to be uniformly integrable $\P$-martingales with respect to $\textbf{F}$ after an appropriate choice of the numéraire. This leads to a Law of One Price. In fact, we have that $P_t=\E_\P\big(P_\infty\,|\,\mathcal{F}_t\big)$ with $P_t=S_t/S^\star_t$ for all $t\geq0$. Since the nominal value process $\big\{S^\star_t\big\}$ is $\textbf{E}$-adapted, we can always substitute $\mathcal{F}_t$ by $\mathcal{E}_t$. Theorem \ref{Th.: Martingale Hypothesis} also clarifies that the DFP $\{\Lambda_t\}$ given by Theorem \ref{Th.: Representation Theorem} is directly related to the GOP. More precisely, $S^\star_t/S^\star_\infty$ represents a \textit{state-price density} or \textit{pricing kernel}, so that
\[
S_t = \E_\P\left(\frac{S^\star_t}{S^\star_\infty}\,S_\infty\,|\,\mathcal{F}_t\right),\qquad\forall~t\geq0.
\]

\citet{Samuelson1965} claims that the nominal price process $\{S_t\}$ is a $\P$-martingale with respect to the natural filtration if future prices are ``properly anticipated.'' In his proof he ignores interest and risk aversion. It is clear that this Martingale Hypothesis cannot be maintained if one takes interest and/or risk preferences into consideration. Theorem \ref{Th.: Martingale Hypothesis} provides a generalization of Samuelson's Martingale Hypothesis. The trick is to apply the ``correct'' discount factor to asset prices, i.e., to choose the GOP as a numéraire, given that the market is complete and sensitive. In this case, we obtain the simple real-world valuation formula $V_t=\E_\P\big(C/S^\star_\infty\,|\,\mathcal{F}_t\big)$ for all $t\geq0$ and each contingent claim $C\in\mathcal{E}_\infty$. This solves the remaining part of the fundamental problem discussed at the beginning of the introduction and at the end of Section \ref{Sec.: The 3rd FTAP}.

The actual challenge is to find the GOP. In practical situations, this can be done by applying econometric procedures. For this purpose, it is not necessary to propagate any specific market model. Another possibility is to approximate the GOP by a linear combination of the market portfolio and the money-market account \citep[][Ch.\ 11]{Platen2006,PH2006}. In either case, since the market is sensitive, it is not necessary to investigate any information that goes beyond the evolution of asset prices and does not exceed the general information flow $\textbf{F}$.

\section{Conclusion}

After an appropriate choice of the numéraire, the discounted price processes in a complete financial market are uniformly integrable martingales under the real-world measure if and only if the market is sensitive. The given result is model independent, i.e., the underlying probabilistic assumptions are minimal, and it highlights two fundamental axioms of neoclassical finance: (i) The absence of arbitrage opportunities and (ii) informational efficiency. An arbitrage opportunity can be either a free lunch with vanishing risk or a dominant strategy. This particular notion of arbitrage is motivated by the 3\textsuperscript{rd} FTAP. Informational efficiency means that the evolution of asset prices is immersed in a general flow of information with respect to the physical measure. Roughly speaking, the market prices must ``fully reflect'' or ``rapidly adjust to'' all relevant information.

To the best of my knowledge this work presents novel results. For example, it extends the 3\textsuperscript{rd} FTAP to markets with infinite lifetime. Further, it illustrates how no-arbitrage conditions, completeness, efficiency, and the growth-optimal portfolio are connected to each other. The presented theorems strengthen the general findings that have been thoroughly discussed in the literature under the label of ``benchmark approach,'' which leads to a Law of Minimal Price. A key observation of this work is that in a complete and sensitive market, the growth-optimal portfolio is determined by the evolution of asset prices and so we obtain a Law of One Price.

The given results could be used for constructing hypothesis tests for market efficiency. For example, one can test the null hypothesis that a market is efficient with respect to the flow of public or private information. Additionally, it is possible to test whether a trader makes use of information that is not ``fully reflected'' by the asset prices. The econometric implications of the presented results and their empirical implementation shall be addressed in the future.

\section*{Acknowledgments}
\addcontentsline{toc}{section}{Acknowledgments}

I would like to thank very much Martin Larsson for his splendid answers to my questions related to martingale theory and his comments on the manuscript. Many thanks belong also to Dirk Becherer, Robert Jarrow, Christoph Memmel, Ilya Molchanov, and Alexander Szimayer.

\begin{appendix}

\section{Appendix}

\subsection{The Classic Approach to Market Efficiency}\label{Sec.: The Classic Approach to Market Efficiency}

The literature on the Efficient-Market Hypothesis is overwhelming and even the number of review papers is huge. Here, I give only a very brief overview of the classic approach to market efficiency.\footnote{See \citet{Sewell2011a} for a comprehensive discussion on the history of the Efficient-Market Hypothesis.} This suggests that the market is a \textit{fair game} \citep{Fama1970}:
\begin{enumerate}
\item[(i)] Each asset has a fair equilibrium expected return conditional on $\mathcal{F}_t$ for all $0\leq t\leq T$ and
\item[(ii)] the true expectations conditional on $\mathcal{F}_t$ coincide with the fair equilibrium expected returns given by (i) at every time $t\in\big[0,T\big]$.\footnote{The assumption that asset returns are serially independent or that they follow a random walk is neither necessary nor sufficient for a fair game \citep{CLM1997,LeRoy1973,Lucas1978}.}
\end{enumerate}

Another, more general, interpretation of market efficiency is due to \citet{Jensen1978}:
\begin{quote}
    ``\textsl{A market is efficient with respect to information set $\theta_t$ if it is impossible to make economic profits by trading on the basis of information set $\theta_t$. By economic profits, we mean the risk adjusted returns net of all costs.}''
\end{quote}

Similarly, \citet{TG2004} conclude that,
\begin{quote}
``\textsl{A market is efficient with respect to the information set, $X_t$, search technologies, $S_t$, and forecasting models, $M_t$, if it is impossible to make economic profits by trading on the basis of signals produced from a forecasting model in $M_t$ defined over predictor variables in the information set $X_t$ and selected using a search technology in $S_t$.}''
\end{quote}

For identifying ``economic profits'' we need to define ``equilibrium expected'' or ``risk adjusted'' asset returns. This leads to the following problem \citep{CLM1997}:
\begin{quote}
``\textsl{[\ldots] any test of efficiency must assume an equilibrium model that defines normal security returns. If efficiency is rejected, this could be because the market is truly inefficient or because an incorrect equilibrium model has been assumed. This \emph{joint hypothesis problem} means that market efficiency as such can never be rejected.}''
\end{quote}

The joint-hypothesis problem can be considered \emph{the} Achilles heel of the classic approach to market efficiency \citep{Fama1991}. There exist many definitions or interpretations of the Efficient-Market Hypothesis. Some of them are discussed in Section \ref{Sec.: Sensitivity}. Definition \ref{Def.: Efficient market} does not require any equilibrium model and thus it is not affected by the joint-hypothesis problem.

\subsection{Arbitrage-Free Markets}

\subsubsection{No-Arbitrage Conditions}\label{Sec.: No-Arbitrage Conditions}

In the following, I use the shorthand notation $\int H\,\d P=\int_0^\infty H_t\,\d P_t$ for the final gain of the strategy $\{H_t\}$.\footnote{Here, it is implicitly assumed that the limit $\int_0^\infty H_t\,\d P_t$ exists almost surely.} An admissible $\textbf{F}$-predictable strategy $\{H_t\}$ that is such that
\begin{enumerate}
  \item[(i)] $\P\left(\int H\,\d P\geq0\right)=1$ and
  \item[(ii)] $\P\left(\int H\,\d P>0\right)>0$
\end{enumerate}
is said to be an \textit{arbitrage}. Now, consider two admissible $\textbf{F}$-predictable strategies $\{G_t\}$ and $\{H_t\}$. The strategy $\{G_t\}$ is said to \textit{dominate} $\{H_t\}$ if and only if
\begin{enumerate}
  \item[(i)] $\P\left(\int G\,\d P\geq \int H\,\d P\right)=1$ and
  \item[(ii)] $\P\left(\int G\,\d P > \int H\,\d P\right)>0\,$.
\end{enumerate}
By contrast, if there is no admissible $\textbf{F}$-predictable strategy that dominates $\{H_t\}$, the latter is said to be \textit{$\textbf{F}$-maximal} \citep{DS1998}.

Dominance can be interpreted as ``relative arbitrage'' \citep{Merton1973}.\footnote{For a similar concept see, e.g., \citet{KF2005} as well as \citet{Platen2004}.} A strategy that is dominated by another strategy can be considered Pareto inefficient. This is because the final gain of the dominating strategy can never be worse, but it is better in some possible states of the world. I say that there is \textit{no dominance} if and only if each single asset in $A$ is $\textbf{F}$-maximal. ND implies \textit{no arbitrage} (NA) but the converse is not true. Moreover, the ND condition implies that no asset can be dominated on \emph{any} time interval $[s,t]$ with $0\leq s<t<\infty\,$. Otherwise, one could hold the corresponding asset from time $0$ to time $s$, switch to the dominant strategy at time $s$, apply this strategy from time $s$ to time $t$, switch back to the asset at time $t$ and maintain this position until the end of time. This would dominate the asset and so the ND condition would be violated.\footnote{See \citet{JL2012} for a similar argument.}

Let $\bigl\{H_{tn}\bigr\}_{n\in\N}$ be a sequence of admissible $\textbf{F}$-predictable strategies and $\int H_n\,\d P$ the final gain of the $n$-th strategy ($n\in\N$).\footnote{Each $\big\{\int_0^tH_{sn}\,\d P_s\big\}$ is bounded below by a \emph{common} number $-a\leq0$. This is implicit in the definition of $K_0$ in \citet[][p.~473]{DS1994}.} The sequence $\{H_{tn}\}$ is said to be a \textit{free lunch with vanishing risk} if and only if there exist some real numbers $\delta,\varepsilon>0$ such that $\big\|(\int H_n\,\d P)^-\big\|_\infty\rightarrow0$ as $n\rightarrow\infty$ and for each $n\in\N$ there exists a natural number $m\geq n$ such that
\[
\P\left(\int H_m\,\d P>\varepsilon\right) > \delta.\footnote{Actually, according to \citet[][Corollary 3.7]{DS1994}, this is a \emph{characterization} of a free lunch with vanishing risk rather than the original definition which is given in topological terms.}
\]
A free lunch with vanishing risk is \emph{essentially} an arbitrage, since the maximum loss can be made arbitrarily small by choosing a sufficiently large $n\in\N$.\footnote{It is worth emphasizing that the loss $\big(\int H_n\,\d P\big)^-$ vanishes \emph{uniformly} (on the essential part of $\Omega$) as $n\rightarrow\infty$ and not only in probability \citep[][p.~501]{DS1994}.} \textit{No free lunch with vanishing risk} implies NA but the converse is not true. NFLVR also guarantees that the final gain $\int H\,\d P$ of every admissible strategy $\{H_t\}$ exists and is finite \citep{DS1994}.

Let $\{H_{tn}\}$ be a sequence of normalized $\textbf{F}$-predictable strategies and $\{V_{tn}\}$ the corresponding sequence of discounted value processes. The sequence $\{H_{tn}\}$ is said to be an \textit{unbounded profit with bounded risk} if and only if $\big\{V_{\infty n}\big\}$ is unbounded in probability, i.e.,
\[
\lim_{x\rightarrow\infty}\,\sup_{n\in\N}\,\P\big(V_{\infty n}>x\big) > 0\,.
\]
This is also referred to as an \textit{arbitrage of the first kind} \citep{IP2011a}. \citet{KK2007} mention that an unbounded profit with bounded risk gives an investor the possibility of making a considerable amount of money out of almost nothing. In fact, there always exist a probability $p>0$ and a real number $x_0>0$ such that $\sup_{n\in\N}\P\big(V_{\infty n}/x>1\big)\geq p$ for all $x\geq x_0\,$. According to \citet[][Proposition 4.2]{KK2007}, there is NUPBR and NA if and only if there is NFLVR, i.e., NFLVR $\Leftrightarrow$ NA $\wedge$ NUPBR.

\subsubsection{Radon-Nikodym Derivatives}\label{Sec.: Radon-Nikodym Derivatives}

For every probability measure $\Q\sim\P$ the Radon-Nikodym Theorem guarantees that there exists one and only one positive $\textbf{F}$-adapted stochastic process $\{\Lambda_t\}$ such that
\[
\int_{F_t} \d\Q = \int_{F_t} \Lambda_t\,\d\P\,,\qquad\forall~F_t\in\mathcal{F}_t\,,~t\geq0\,.
\]
The random variable $\Lambda_t$ represents the Radon-Nikodym derivative of $\Q$ with respect to $\P$ on the $\sigma$-algebra $\mathcal{F}_t$. Therefore, $\{\Lambda_t\}$ is referred to as the \textit{Radon-Nikodym process} with respect to the filtration $\textbf{F}$ associated with $\Q\,$. Moreover, there exists one and only one positive random variable $\Lambda$ such that $\int_F \d\Q = \int_F \Lambda\,\d\P$ for all $F\in\mathcal{F}_\infty$ and thus
\[
\int_{F_t} \Lambda_t\,\d\P = \int_{F_t} \Lambda\,\d\P\,,\qquad\forall~F_t\in\mathcal{F}_t\,,~t\geq0\,.
\]
This means it holds that $\Lambda_t=\E_\P\bigl(\Lambda\,|\,\mathcal{F}_t\bigr)$ for all $t\geq0$ and so $\{\Lambda_t\}$ is a uniformly integrable $\P$-martingale with respect to $\textbf{F}$ converging to $\Lambda>0\,$.\footnote{Lévy's Zero-One Law leads to $\Lambda_t=\E_\P(\Lambda\,|\,\mathcal{F}_t)\rightarrow\E_\P(\Lambda\,|\,\mathcal{F}_\infty)=\Lambda$ as $t\rightarrow\infty\,$.} The inverse RNP $\bigl\{\Lambda^{-1}_t\bigr\}$ with respect to $\textbf{F}$ carries the Radon-Nikodym derivatives of $\P$ with respect to $\Q$ and is a uniformly integrable $\Q$-martingale with respect to $\textbf{F}$ converging to $\Lambda^{-1}>0\,$.

\begin{lem}\label{Lem.: Change of measure}
Consider a probability measure $\Q\sim\P$. Let $\{\Lambda_t\}$ be the associated $\textbf{F}$-RNP and fix any $T<\infty\,$. Then for every random variable $X\in\mathcal{F}_T$ with $\E_\Q(|X|)<\infty$ we obtain
\[
\E_\Q(X\,|\,\mathcal{F}_t) = \E_\P\left(\frac{\Lambda_T}{\Lambda_t}\,X\,|\,\mathcal{F}_t\right)
\]
for all $0\leq t\leq T$. Moreover, for every random variable $X\in\mathcal{F}_\infty$ with $\E_\Q(|X|)<\infty$ we have that
\[
\E_\Q(X\,|\,\mathcal{F}_t) = \E_\P\left(\frac{\Lambda_\infty}{\Lambda_t}\,X\,|\,\mathcal{F}_t\right)
\]
for all $t\geq0\,$.
\end{lem}

Proof: Since $\E_\Q(|X|)<\infty$ and $X$ as well as $\Lambda_T$ are $\mathcal{F}_T$-measurable, we have that
\begin{align*}
\int_{F_t}X\,\d\Q &= \int_{F_t}\Lambda_TX\,\d\P = \int_{F_t}\E_\P(\Lambda_TX\,|\,\mathcal{F}_t)\,\d\P \\
&= \int_{F_t}\E_\P\left(\frac{\Lambda_T}{\Lambda_t}\,X\,|\,\mathcal{F}_t\right)\Lambda_t\,\d\P = \int_{F_t}\E_\P\left(\frac{\Lambda_T}{\Lambda_t}\,X\,|\,\mathcal{F}_t\right)\d\Q
\end{align*}
for all $0\leq t\leq T$ and $F_t\in\mathcal{F}_t\,$, i.e., $\E_\Q(X\,|\,\mathcal{F}_t)=\E_\P\big[(\Lambda_T/\Lambda_t)\,X\,|\,\mathcal{F}_t\big]$. Similar arguments apply in the case $X\in\mathcal{F}_\infty$.\QED

Due to Lemma \ref{Lem.: Change of measure}, we have that
\[
\E_\Q(X\,|\,\mathcal{F}_t) = \E_\P\left(\frac{\Lambda_T}{\Lambda_t}\,X\,|\,\mathcal{F}_t\right) = \E_\P\left(\frac{\Lambda_T/\Lambda_0}{\Lambda_t/\Lambda_0}\,X\,|\,\mathcal{F}_t\right)
\]
for all $0\leq t\leq T<\infty\,$. Hence, we can focus on the \emph{normalized} RNP $\{\Lambda_t/\Lambda_0\}$. Throughout this work it is implicitly assumed that $\Lambda_0=1$ without loss of generality.

\begin{lem}\label{Lem.: Change of martingale}
Consider a probability measure\/ $\Q\sim\P$ and let $\{\Lambda_t\}$ be the associated $\textbf{F}$-RNP. The stochastic process\/ $\{X_t\}$ is a (local)\/ $\Q$-martingale with respect to $\textbf{F}$ if and only if\/ $\{\Lambda_tX_t\}$ is a (local)\/ $\P$-martingale with respect to $\textbf{F}$.
\end{lem}

Proof: For the ``only if'' part suppose that $\{X_t\}$ is a local $\Q$-martingale with respect to $\textbf{F}$ and consider a localizing sequence $\{\tau_n\}$ of $\textbf{F}$-stopping times, so that
\[
X_{t\wedge\tau_n} = \E_\Q(X_{T\wedge\tau_n}\,|\,\mathcal{F}_t) = \E_\P\left(\frac{\Lambda_T}{\Lambda_t}X_{T\wedge\tau_n}\,|\,\mathcal{F}_t\right),\qquad\forall~n\in\N\,,~0\leq t\leq T<\infty\,.
\]
Hence, the stochastic process $\{\Lambda_tX_{t\wedge\tau_n}\}$ is a $\P$-martingale with respect to $\textbf{F}$ for all $n\in\N\,$. It follows that $\{\Lambda_{t\wedge\tau_n}X_{t\wedge\tau_n}\}$ is also a $\P$-martingale with respect to $\textbf{F}$ for all $n\in\N$.\footnote{More precisely, $\{\Lambda_{t\wedge\tau_n}X_{t\wedge\tau_n}\}$ is obtained by stopping $\{\Lambda_tX_{t\wedge\tau_n}\}$ once again by $\{\tau_n\}$.} This means $\{\Lambda_tX_t\}$ is a local $\P$-martingale with respect to $\textbf{F}$. For the ``if'' part suppose that $\{Y_t\}$ with $Y_t=\Lambda_tX_t$ for all $t\geq0$ is a local $\P$-martingale with respect to $\mathcal{\textbf{F}}$. Now, there exists a localizing sequence $\{\tau_t\}$, so that
\[
Y_{t\wedge\tau_n} = \E_\P(Y_{T\wedge\tau_n}\,|\,\mathcal{F}_t) = \E_\Q\left(\frac{\Lambda^{-1}_T}{\Lambda^{-1}_t}X_{T\wedge\tau_n}\,|\,\mathcal{F}_t\right),\qquad\forall\,n\in\N\,,~0\leq t\leq T<\infty\,.
\]
This means $\bigl\{\Lambda^{-1}_{t\wedge\tau_n}Y_{t\wedge\tau_n}\bigr\}$ is a $\Q$-martingale with respect to $\textbf{F}$. Since $\Lambda^{-1}_{t\wedge\tau_n}Y_{t\wedge\tau_n}=X_{t\wedge\tau_n}$ for all $t\geq0\,$, $\{X_t\}$ is a local $\Q$-martingale with respect to $\textbf{F}$. Similar arguments apply without localization if $\{X_t\}$ is a $\Q$-martingale or $\{\Lambda_tX_t\}$ is a $\P$-martingale with respect to \textbf{F}.\QED

\begin{lem}\label{Lem.: RNP <-> DFP}
Let $a\in\mathcal{A}$ be some numéraire asset and suppose that $\mathcal{M}^a(\textbf{F})\neq\emptyset$ ($\mathcal{L}^a(\textbf{F})\neq\emptyset$). Then the $\textbf{F}$-RNP associated with $\Q\in\mathcal{M}^a(\textbf{F})$ ($\Q\in\mathcal{L}^a(\textbf{F})$) is a (local) $\textbf{F}$-DFP associated with $\Q$ and a (local) $\textbf{F}$-DFP associated with $\Q$ is the $\textbf{F}$-RNP associated with $\Q\in\mathcal{M}^a(\textbf{F})$ ($\Q\in\mathcal{L}^a(\textbf{F})$).
\end{lem}

Proof: The $\textbf{F}$-RNP $\{\Lambda_t\}$ associated with $\Q\in\mathcal{L}^a(\textbf{F})$ is a positive uniformly integrable $\P$-martingale with respect to $\textbf{F}$ such that $\Lambda_0=1$ and $\Lambda_\infty>0\,$. Every discounted price process $\{P_t\}$ is a local $\Q$-martingale with respect to $\textbf{F}$ and Lemma \ref{Lem.: Change of martingale} implies that $\{\Lambda_tP_t\}$ is a local $\P$-martingale with respect to $\textbf{F}$. Hence, $\{\Lambda_t\}$ is a local $\textbf{F}$-DFP associated with $\Q\,$. Conversely, let $\{\Lambda_t\}$ be a local $\textbf{F}$-DFP associated with $\Q\,$, i.e.,
\[
\Q(F) = \int_F\Lambda_\infty\,\d\P\,,\qquad\forall\,F\in\mathcal{F}_\infty\,.
\]
It holds that
\[
\Q(F_t) = \int_{F_t}\Lambda_\infty\,\d\P = \int_{F_t}\E_\P(\Lambda_\infty\,|\,\mathcal{F}_t)\,\d\P = \int_{F_t}\Lambda_t\,\d\P\,,\qquad\forall\,F_t\in\mathcal{F}_t\,,~t\geq0\,.
\]
Since $\Lambda_0=1$ and $\Lambda_t>0$ for all $t\geq0\,$, the Radon-Nikodym Theorem guarantees that $\Lambda_t$ is the Radon-Nikodym derivative on the $\sigma$-algebra $\mathcal{F}_t$ for all $t\geq0\,$. Since $\Lambda_\infty>0$ we have that $\Q\sim\P$. Moreover, $\{\Lambda_t\}$ leads to a local $\P$-martingale $\{\Lambda_tP_t\}$ with respect to $\textbf{F}$. From Lemma \ref{Lem.: Change of martingale} it follows that $\{P_t\}$ is a local $\Q$-martingale with respect to $\textbf{F}$ and thus $\Q\in\mathcal{L}^a(\textbf{F})$. Similar arguments apply if $\Q\in\mathcal{M}^a(\textbf{F})$ or $\{\Lambda_t\}$ is an $\textbf{F}$-DFP, respectively.\QED

\begin{prop}\label{Pr.: Local martingale deflator}
Let $a\in\mathcal{A}$ be some numéraire asset and $\{V_t\}$ the discounted value process of any admissible $\textbf{F}$-predictable strategy. Suppose that there exists a local $\textbf{F}$-DFP $\{\Lambda_t\}$. Then $\{\Lambda_tV_t\}$ is a local\/ $\P$-martingale with respect to $\textbf{F}$.
\end{prop}

Proof: Let $\Q$ be the ELMM associated with $\{\Lambda_t\}$. According to Lemma \ref{Lem.: RNP <-> DFP}, $\{\Lambda_t\}$ represents the $\textbf{F}$-RNP associated with $\Q\in\mathcal{L}^a(\textbf{F})$. From the Ansel-Stricker Theorem \citep{AS1994} we conclude that $\{V_t\}$ is a local $\Q$-martingale with respect to $\textbf{F}$. Now, Lemma \ref{Lem.: Change of martingale} implies that $\{\Lambda_tV_t\}$ is a local $\P$-martingale with respect to $\textbf{F}$.\QED

The last result shows that $\{\Lambda_t\}$ is a local martingale deflator and as such it is also an equivalent supermartingale deflator \citep{JL2013,KK2007}. This reminds us of the supermartingale property of the benchmarked value process $\big\{W^{-1}_tV_t\big\}$, where $\{W_t\}$ is the discounted value process of the GOP with respect to $\textbf{F}$. In general, the inverse of any equivalent supermartingale deflator $\{\Lambda_t\}$ need not be the discounted value process of the GOP or any other (1-)admissible trading strategy. Hence, $\big\{\Lambda^{-1}_t\big\}$ is not necessarily a \emph{tradeable} supermartingale deflator \citep{KKS2015,KK2007}. Nonetheless, from Theorem 4.12 in \citet{KK2007} we conclude that there exists a tradeable supermartingale deflator if and only if there is NUPBR. Moreover, there is NUPBR if and only if \emph{any} equivalent supermartingale deflator exists.

\subsubsection{Stochastic Discount Factors}

Stochastic discount factors are frequently used in finance literature \citep{Cochrane2005}. In the following, $\{\Pi_t\}$ denotes the scalar-valued discounted price process of an arbitrary asset. The basic pricing formula
\[
\Pi_t = \E_\P\left(\Lambda_{t,T}\Pi_T\,|\,\mathcal{F}_t\right) = \E_\Q\bigl(\Pi_T\,|\,\mathcal{F}_t\bigr)
\]
for all $0\leq t\leq T<\infty$, where $\{\Lambda_t\}$ is an $\textbf{F}$-DFP, implies that
\[
\E_\P\left[\Lambda_{t,T}\left(\frac{\Pi_T}{\Pi_t}-1\right)|\,\mathcal{I}_t\right] = \E_\Q\left(\frac{\Pi_T}{\Pi_t}-1|\,\mathcal{I}_t\right) = 0
\]
for every $\sigma$-algebra $\mathcal{I}_t\subseteq\mathcal{F}_t$ and all $0\leq t\leq T<\infty$. Here, $\Q$ denotes the EMM with respect to $\textbf{F}$ associated with $\{\Lambda_t\}$ and $\Pi_T/\Pi_t-1$ is the return on the given asset between time $t$ and $T$. This means future asset returns cannot be predicted under the EMM $\Q$ on the basis of any information set $\mathcal{I}_t\subseteq\mathcal{F}_t$, but under the physical measure $\P$, they are possibly predictable. Hence, contrary to common belief, market efficiency does not rule out predictability of asset returns \citep{TG2004}.

If $\{\Lambda_t\}$ is a \emph{local} $\textbf{F}$-DFP it can only be guaranteed that
\[
\E_\P\left[\Lambda_{t,T}\left(\frac{\Pi_T}{\Pi_t}-1\right)|\,\mathcal{I}_t\right] = \E_\Q\left(\frac{\Pi_T}{\Pi_t}-1|\,\mathcal{I}_t\right) \leq 0
\]
for every $\sigma$-algebra $\mathcal{I}_t\subseteq\mathcal{F}_t$ and all $0\leq t\leq T<\infty\,$. Nevertheless, the expected asset return conditional on $\mathcal{I}_t$ still might be positive under $\P$.

Once again, let $\{\Lambda_t\}$ be an $\textbf{F}$-DFP. An important feature of the basic pricing formula is that
\[
\E_\P\left(\Lambda_{t,T}\Pi_T\,|\,\mathcal{E}_t\right) = \E_\P\Big[\E_\P\bigl(\Lambda_{t,T}\Pi_T\,|\,\mathcal{F}_t\bigr)\,|\,\mathcal{E}_t\Big] = \E_\P\bigl(\Pi_t\,|\,\mathcal{E}_t\bigr) = \Pi_t
\]
for all $0\leq t\leq T<\infty\,$. Hence, stochastic discount factors are ``downward compatible,'' i.e., every discount factor that has been calculated on the basis of $\textbf{F}$ can be applied to $\textbf{E}$.\footnote{In general, the $\textbf{F}$-DFP $\{\Lambda_t\}$ is not $\textbf{E}$-adapted and thus, although we have that $\E_\P(\Lambda_{t,T}\Pi_T\,|\,\mathcal{E}_t)=\Pi_t$ for all $0\leq t\leq T<\infty\,$, downward compatibility does not imply that $\{\Lambda_t\Pi_t\}$ is a $\P$-martingale with respect to $\textbf{E}$.}

More precisely, let $\bigl\{\Lambda^{\mathcal{F}}_t\bigr\}$ be an $\textbf{F}$-DFP. Each discount factor $\Lambda^{\mathcal{F}}_{t,T}$ is $\mathcal{F}_T$-measurable but not necessarily $\mathcal{E}_T$-measurable. Nevertheless, since $\mathcal{M}(\textbf{E})\supseteq\mathcal{M}(\textbf{F})=\emptyset$, there also exists an $\textbf{E}$-DFP $\bigl\{\Lambda^{\mathcal{E}}_t\bigr\}$, so that
\[
\Pi_t = \E_\P\left(\Lambda^{\mathcal{F}}_{t,T}\Pi_T\,|\,\mathcal{E}_t\right) = \E_\P\left(\Lambda^{\mathcal{E}}_{t,T}\Pi_T\,|\,\mathcal{E}_t\right),
\]
but in general
\[
\E_\P\left(\Lambda^{\mathcal{E}}_{t,T}\Pi_T\,|\,\mathcal{E}_t\right) \neq \E_\P\left(\Lambda^{\mathcal{E}}_{t,T}\Pi_T\,|\,\mathcal{F}_t\right)
\]
for any $0\leq t\leq T<\infty\,$. Put another way, stochastic discount factors are not ``upward compatible,'' which means that a discount factor that has been calculated on the basis of $\textbf{E}$ in general cannot be applied to a broader filtration $\textbf{F}$. Hence, if somebody aims at representing asset prices with respect to the information set $\mathcal{F}_t\,$, he or she must use a discount factor that is made for $\textbf{F}$ or for any superfiltration $\textbf{G}\supset\textbf{F}$. This is cumbersome or even impossible in most practical situations.

For this reason, it is highly desirable to know under which circumstances stochastic discount factors that have been calculated on the basis of $\textbf{E}$ can be applied also to a broader filtration $\textbf{F}$. In fact, this is the key property of a sensitive market, since market sensitivity guarantees that
\[
\E_\P\left(\Lambda^{\mathcal{E}}_{t,T}\Pi_T\,|\,\mathcal{E}_t\right) = \E_\P\left(\Lambda^{\mathcal{E}}_{t,T}\Pi_T\,|\,\mathcal{F}_t\right)
\]
for all $0\leq t\leq T<\infty\,$. Put another way, if the market is sensitive, the discount factor $\Lambda^{\mathcal{E}}_{t,T}$ contains all relevant information. 

\end{appendix}

\bibliographystyle{C:/Cloud/Local/LaTeX/BibTeX/mybib}
\bibliography{C:/Cloud/Local/LaTeX/BibTeX/mybib}

\end{document}